\documentclass[11pt]{article}

\usepackage{graphicx}
\usepackage{mathrsfs}
\usepackage{latexsym}
\usepackage{amsmath}
\usepackage{amssymb}
\usepackage{color}

\usepackage[nohead,letterpaper]{geometry}
\usepackage{vmargin}
\setpapersize{USletter}
\setmargrb{2cm}{1cm}{2cm}{1.75cm}

\usepackage{xcolor}
\definecolor{plum}{rgb}{0.36078, 0.20784, 0.4}
\definecolor{chameleon}{rgb}{0.30588, 0.60392, 0.023529}
\definecolor{cornflower}{rgb}{0.12549, 0.29020, 0.52941}
\definecolor{scarlet}{rgb}{0.8, 0, 0}
\definecolor{brick}{rgb}{0.64314, 0, 0}

\newcommand{\email}[1]{\href{mailto:#1}{\tt \textcolor{cornflower}{#1}}}

\newcommand{\ba}{\begin{eqnarray}}
\newcommand{\ea}{\end{eqnarray}}
\newcommand{\be}{\begin{equation}}
\newcommand{\ee}{\end{equation}}
\newcommand{\bd}{\begin{displaymath}}
\newcommand{\ed}{\end{displaymath}}

\usepackage[%
  pdftitle={Holographic Renormalization of Asymptotically Flat gravity},%
  pdfauthor={Miok Park, Robert Mann},%
  pdfsubject={Holographic Renormalization of Asymptotically Flat gravity},%
  pdfkeywords={Holographic Renormalization, Boundary Terms, Surface Terms, Variational
    Principle},%
  colorlinks=true,linkcolor=cornflower,citecolor=brick,urlcolor=chameleon%
  ]{hyperref}

\numberwithin{equation}{section}

\usepackage[bf,medium,]{titlesec}
\usepackage{color}

\titleformat{\section}
{\large\bfseries}{\thesection.}{.5em}{\large\bfseries}

\titleformat{\subsection}
{\bfseries}{\thesubsection}{.5em}{\bfseries}

\begin{document}

%%%%%%%%%%%%%%%%%%%%%%%%%%%%%%%%%%%%%%%%%%%%
% Titlepage
%%%%%%%%%%%%%%%%%%%%%%%%%%%%%%%%%%%%%%%%%%%%

\thispagestyle{empty}
%\preprint{XXXX}

%\begin{flushright}
%\today
%\end{flushright}
~\vspace{2cm}\\

\begin{center}
{\bf \Large Holographic Renormalization of Asymptotically Flat Gravity}
\end{center}

\vspace{.5cm}

\begin{center}
Miok Park $^{a}$ and Robert B. Mann $^{a, b}$

\vspace{1cm}{\small {\textit{$^{a}$Department of Physics,\\ University of
Waterloo,\\ Waterloo, Ontario N2L 3G1,\\ Canada}}}\\
\vspace{2mm} {\small {\textit{$^{b}$Perimeter Institute for Theoretical Physics,\\ 31 Caroline Street North,\\ Waterloo,
Ontario N2L 2Y5,\\ Canada}}}\\
\vspace*{0.5cm}
\email{m7park@sciborg.uwaterloo.ca}\,,
\email{rbmann@sciborg.uwaterloo.ca}
\end{center}
\vspace{0.5cm}

\begin{center}
{\bf Abstract}
\end{center}
We compute the boundary stress tensor associated with Mann-Marolf counterterm in asymptotic flat and static spacetime for cylindrical boundary surface as $r \rightarrow \infty$, and find that the form of the boundary stress tensor is the same as the hyperbolic boundary case in 4 dimensions, but has additional terms in higher than 4 dimensions. We find that these additional terms are impotent and do not contribute to conserved charges. We also check the conservation of the boundary stress tensor in a sense that $\mathcal{D}^a T_{ab} = 0$, and apply our result to the ($n+3$)-dimensional static black hole solution. As a result, we show that the stress boundary tensor with Mann-Marolf counterterm works well in standard boundary surfaces.

\newpage

\setcounter{page}{1}

%%%%%%%%%%%%%%%%%%%%%%%%%%%%%%%%%%%%%%%%%%%%
% Table of Contents
%%%%%%%%%%%%%%%%%%%%%%%%%%%%%%%%%%%%%%%%%%%%
\tableofcontents

%%%%%%%%%%%%%%%%%%%%%%%%%%%%%%%%%%%%%%%%%%%%
% Body of Paper
%%%%%%%%%%%%%%%%%%%%%%%%%%%%%%%%%%%%%%%%%%%%

%----------------------------------------------
%----------------------------------------------
\section{Introduction}
\label{sec:Intro}
%----------------------------------------------
%----------------------------------------------

The asymptotic structure of  gravitating systems has been of interest for many reasons since the advent of relativity. The original aim is to understand the basic properties of the space-time in terms of its relationship to global quantities like total energy, momentum, and angular momentum. Another reason arises in numerical relativity, whose main purpose  is to study  time evolution of gravitational systems; hence setting  proper boundary conditions and imposing initial data are of crucial import. Related to the former, one knows that a system is described and characterized by locally defined tensor fields. In special relativity, it is easy to figure out the global characteristics of the system, because vector addition holds on the manifold. However, if curvature exists then it is problematic  to collect together all local contributions of a given quantity. One approach for addressing this problem in general relativity is to study a given region far away from the source. In that asymptotic region, the curvature becomes much weaker (assuming cosmological effects are subdominant) and the spacetime is very close to Minkowski space. One could then expect to define global quantities for the whole system. However there still exists a serious ambiguity in exploring the asymptotic structure because of various ways of approaching infinity, and defining or classifying the notion of asymptotically flat spacetimes.

Early work associated with an asymptotic geometry originated with from Arnowitt, Deser and Misner (ADM),  who produced an integral expression for  an asymptotically flat spacetime's energy-momentum and angular momentum via a $3+1$ decomposition, which slices the spacetime into space and time {\cite{Arnowitt:1960}}.  This approach yielded conserved (ADM) quantities at spatial infinity, $r \rightarrow \infty$. Geroch later reformulated and extended this work by providing a definition of asymptotic flatness at "spatial infinity" {\cite{Geroch:1972}}. These $(3+1)$ frameworks, however, have difficulties encompassing the concept of an asymptotic spacetime and its symmetry group at near infinity {\cite{Ashtekar:1992}}.

Shortly after ADM, Bondi et al. {\cite{Bondi:1962}} considered an isolated body emitting radiation such as a scalar field, an electromagnetic wave, or a gravitational wave propagating to null infinity instead of spatial infinity and obtained the Bondi-Sachs mass. For a stationary spacetime, it was proved that the ADM 4-momentum at spatial infinity is the past limit of the Bondi-Sachs 4 momentum {\cite{Ashtekar:1979}}. Based on this work, Penrose elegantly formulated a definition of an asymptotically flat spacetime by introducing the concept of future and past "null infinity" $(\mathfrak{I}^{+}, \mathfrak{I}^{-})$, using a conformal completion method {\cite{Penrose:1963}}. In this picture, for Minkowski space, $\mathfrak{I}^{+}$ and $\mathfrak{I}^{-}$ meet at spacelike infinity ($r \rightarrow \infty$ at fixed $t$), which can be described by a point $I^{0}$ in the conformal extension of Minkowski space. For a curved static spacetime, e.g. the Schwarzschild metric, all points at spatial infinity are squeezed down to a single point, so the point $I^{0}$ loses some essential properties that it has in the flat case {\cite{Beig:1982}}.

These two notions of asymptotic flatness at null and spatial infinity were unified into a single notion by Ashtekar and Hansen in {\cite{Ashtekar:1978}}. They formulated asymptotic conditions that treated spacetime as a whole rather than splitting space and time and forged a link between the asymptotic symmetry group and conserved quantities.

The importance of considering a spacetime boundary  has also been of interest in terms of the gravitational action. From the existing Einstein-Hilbert action, Gibbons and Hawking pointed out that variations of metric derivatives at the boundary must not be ignored, and introduced  the Gibbons-Hawking boundary term to fix this problem,
\begin{equation}
{S_{EH + GH} =  \frac{1}{16 \pi G} \int_{\mathcal{M}} \sqrt{-g} R + \frac{1}{8 \pi G} \int_{\partial \mathcal{M}} \sqrt{-h} K \label{EGactn}}
\end{equation}
where $g$ is a trace of the metric on the spacetime, $R$ is a Ricci scalar with respected to $g_{ab}$, $h$ is the trace of the induced metric on the boundary, and $K$ is the trace of its extrinsic curvature.  While this new boundary term is well defined for spatially compact spacetimes, it diverges for noncompact ones.  Remedies for this problem have involved adding a non-dynamical term into the action. Two main ideas on how to characterize this term have been suggested.

The first idea is the reference background method,  originally suggested by Brown and York \cite{Brown:1992br}  for asymptotically flat spacetimes and extended to asymptotically anti de Sitter spacetimes by Brown, Creighton and Mann \cite{Brown:1994gs}. The basic idea is to introduce a reference background $(g_0, \phi_0)$ (required to be a static solution to the field equations  {\cite{Hawking:1996}}),  and to write the physical action as
 \begin{equation} \label{bstm}
I_{p} (g , \phi) \equiv I (g, \phi) - I(g_{0}, \phi_{0})
\end{equation}
  where  $I_{p}$ is zero for the reference background, and is finite provided that the fields $(g, \phi)$ match the corresponding fields $(g_{0}, \phi_{0})$ on a proper boundary, i.e. near infinity. For an asymptotically flat spacetime with no matter fields, the gravitational action becomes $S_{EH+GH}$ plus the additional term
\begin{equation}
I(g_{0}) = S_{0} = - \frac{1}{8 \pi G} \int_{\partial \mathcal{M}} \sqrt{-h} K_{0}
\end{equation}
where $K_{0}$ is the trace of the extrinsic curvature of the boundary $(\partial \mathcal{M}, h)$, and is determined by taking an appropriate limit and then matching the boundary metric with the reference metric $(\mathcal{M}_{0}, g_{0})$
embedded in the flat reference spacetime. The more general form $I(g_{0}, \phi_{0})$ may have to be used for certain matter fields
\cite{MannRoss:1995}. However such a proper reference spacetime in general does not exist for dimensions larger than 3. This is because  an embedding is required not just  for a particular boundary spacetime, but instead for an open set of boundary spacetimes associated with arbitrary small metric/matter variations. For $d>3$,  given any embeddable boundary spacetime, there are spacetimes arbitrarily nearby that are not  embeddable  {\cite{Mann:2006}}.

The second idea is the counterterm method, whose  form
\begin{equation}\label{KCT}
S_{CT} = - \frac{1}{8 \pi G} \int_{\partial \mathcal{M}} \sqrt{-h} \hat{K}_{CT} (h)
\end{equation}
is added to the action (\ref{EGactn}). The counterterm $\hat{K}_{CT}$ is defined to be a  functional only of geometric invariants of the induced metric $h_{ab}$,  chosen to cancel the divergences in (\ref{EGactn}). Construction of conserved quantities associated with the counterterm method via the renormalized boundary stress tensor was originally developed for an asymptotically anti-de Sitter spacetime  \cite{Balasubramanian:1999re,Mann:1999pc,Emparan:1999pm}.
Motivated by this success, Mann and Marolf extended this method to asymptotically flat spacetime, in which the covariant counterterm
$\hat{K}_{ab}$ is computed from the relation
\begin{equation}
{\cal{R}}_{ab} = \hat{K}_{ab} \hat{K} - \hat{K}_{a}^{\; \; c} \hat{K}_{cb} \label{MMrltn}
\end{equation}
where $\mathcal{R}_{ab}$ is the Ricci tensor of $h_{ab}$ induced on $\partial \mathcal{M}$ and $\hat{K}$ is a trace of $\hat{K}_{ab}$ contracted with $h^{ab}$. The motivation behind eq. (\ref{MMrltn}) is from the Gauss-Codazzi relation
\begin{equation}
\mathcal{R}_{abcd} = R^{Ref}_{abcd} + K_{ac} K_{bd} - K_{ad} K_{bc} \label{GCrlt}
\end{equation}
where $\mathcal{R}_{abcd}$ and $R^{Ref}_{abcd}$ are respectively the Riemann tensor  on $\partial \mathcal{M}$ and on the bulk spacetime $\mathcal{M}$. For an asymptotically flat spacetime $R^{Ref}_{abcd}$ obviously vanishes.  Replacing $K_{ab}$ with
a tensor $\hat{K}_{ab}$ and contracting (\ref{GCrlt}) with $h^{cd}$ yields (\ref{MMrltn}). It has been proven that including the
counter term (\ref{KCT}) leads an action that is finite on asymptotically flat spacetime and stationary under metric variations for two standard asymptotic hypersurfaces, respectively referred to as  "cylindrical" and "hyperbolic" boundary spacetimes $(\partial \mathcal{M} ,h)$ in {\cite{Mann:2006}}.

The boundary stress tensor is defined as the functional derivative of the on-shell action with respect to $h_{ab}$, which takes the form
\begin{equation}
T^{\pi}_{ab} = - \frac{2}{\sqrt{-h}} \frac{\delta S}{\delta h^{ab}} = \frac{1}{8 \pi G} \bigg( \pi_{ab} - \hat{\pi}_{ab} \bigg) \label{BStnsr_pi}
\end{equation}
where $h_{ab}$ is a induced metric on the asymptotic boundary, $\pi_{ab}= K_{ab} - K h_{ab}$ is the conjugate momentum of the gravitational field, and $\hat{\pi}_{ab}$ is an analogous contribution from the counterterm $\hat{K}_{CT} $. Then the conserved charge associated with the Killing vector $\xi^{a}$ via (\ref{BStnsr_pi})  in the cylindrical coordinates  is
\begin{equation}
Q[\xi] = \oint_{C_{r}} d^{n+1} x \sqrt{\gamma_{C_{r}}} u^{a}_{C_{r}} T^{\pi}_{ab} \xi^{a} \label{cnsvdcg}
\end{equation}
in ($n+3$) dimensions, where $\gamma$ is the trace of the induced metric on the $r=const.$ boundary at $t=const.$, and $C_{r}$ is a Cauchy surface within a constant $r$ hypersurface $\mathcal{H}_{r}$ such that $C=\lim_{r \rightarrow \infty} C_{r}$ is a Cauchy surface in the cylindrical boundary $\mathcal{H}$, and $u^{a}$ is a timelike unit vector normal to $C$ in $\mathcal{H}_r$.

In practice, however, the variation of the action has additional terms as a consequence of the definition (\ref{MMrltn}); these are
 represented by $\Delta^{ab}$  and must be added to eq.  (\ref{BStnsr_pi}). Despite this, we shall demonstrate that the quantities $\Delta^{ab}$  do not modify either conserved quantities as given by (\ref{cnsvdcg}) or the conservation of the boundary stress-energy
 for cylindrical boundary conditions.
 Investigation of the  connection between the boundary stress energy  in (\ref{BStnsr_pi}) with the counter term definition (\ref{MMrltn}) indicated that the extra term $\Delta^{ab}$ vanishes  for higher than $4$-dimensional spaceetime and makes no contribution to the conserved charge for 4-dimensional spacetime  \cite{Mann:2008}.  These computations were
 carried out using hyperboloid coordinates for the boundary of the asymptotically flat spacetime,  compatible with the previous studies {\cite{Beig:1982}}, {\cite{Beig:1984}}, and {\cite{Ashtekar:1978}}.   Specifically  the conserved charges were shown to agree \cite{Mann:2008} with   those  defined by Ashtekar and Hansen {\cite{Ashtekar:1978}}.

Here we investigate the boundary stress tensor method (\ref{BStnsr_pi}) associated with Mann-Marolf counterterm for  cylindrical boundary conditions in this paper. As the structure of the boundary and the falloff rates of the metric components differ from those in the hyperbolic case, our aim is to understand the role played by $\Delta^{ab}$ in the context of defining a boundary stress-energy and conserved charges.   As many spacetimes are commonly described in coordinates that asymptote to cylindrical ones,
using the cylindrical boundary condition thus has great practical advantages for computation.  By contrast,  hyperboloid coordinates are rather impractical insofar as they require a non-trivial transformation of the coordinates of most asymptotically flat metrics.

With the same purpose, Astefanesei, Mann, and Stelea  made some preliminary investigations using
 cylindrical coordinate, but considered only leading order  fall-off conditions on components of the metric {\cite{Astefanesei:2007}}.  This is not sufficient for understanding the role played by $\Delta^{ab}$ in the asymptotically flat boundary counterterm approach using
 (\ref{MMrltn}).

We begin with defining an $(n+3)$-dimensional asymptotically flat and static spacetime in cylindrical coordinates, whose metric functions  fall off at least as fast  as $r^{-(n+2)}$.  We then compute $\Delta^{ab}$ as a power series in $1/r$ up to the relevant fall off levels that could potentially affect the conserved charges.  We find that i) to leading order, $\Delta^{ab}$ is manifestly zero for arbitrary dimensions, ii) the first sub-leading order of $\Delta^{ab}$ for $n=1$  is zero, iii) $\Delta^{ab}$ does not vanish for $n>1$, but iv) its non-vanishing does not affect  conserved quantities. In addition, we find that for $n \geq 2$ manifestly ${\mathcal{D}}^a T_{ab} = 0$, but for $n=1$ satisfying ${\mathcal{D}}^a T_{ab} = 0$ requires a condition between higher-order coefficients in the metric, and this calculation is described in appendix F. Finally, we show explicitly how the conserved quantity formula (\ref{cnsvdcg}) associated with the counterterm (\ref{MMrltn}) works in $(n+3)$-dimensional static spacetime.

Our paper proceeds as follows. In section 2, we review a variation of the action and the boundary stress tensor demonstrated already in {\cite{Mann:2006}}, {\cite{Mann:2008}}, and introduce our definition of  asymptotic flatness in the cylindrical coordinates. Section 3 explains the process of deriving $\Delta^{ab}$ and exhibits its explicit form: first the extrinsic curvature, $K_{ab}$, of the asymptotic boundary (with normal vector, $n^{\alpha}$) is calculated, and the result is inserted  into the decomposed Einstein equations. Once the boundary surface Ricci tensor, $\mathcal{R}_{ab}$,  is obtained, the counterterm $\hat{K}_{ab}$ via (\ref{MMrltn}) can be found. From this  $\Delta^{ab}$ is eventually computed in terms of $\hat{K}_{ab}$. As it turns out that $\Delta^{ab}$ is not zero, we investigate the how it is related to the conserved quantity formula (\ref{cnsvdcg}), and show that  $\mathcal{D}_{a} T^{ab} = 0$. In section 4, we provide explicit examples of how to compute conserved charges in $(n+3)$-dimensional static spacetime.

%----------------------------------------------
%----------------------------------------------
\section{Preliminaries}
\label{Cyn}
%----------------------------------------------
%----------------------------------------------

%----------------------------------------------
\subsection{A Variation of the action and the Boundary Stress Tensor}
\label{sec:EOM1}
%----------------------------------------------

The action we start with is
\begin{equation}
S = \frac{1}{16 \pi G} \int_{{\cal{M}}} \sqrt{-g} R + \frac{1}{8 \pi G} \int_{\partial {\cal{M}}} \sqrt{-h} (K - \hat{K}) \label{ACTwtCC}
\end{equation}
where the first and second term are the Einstein-Hilbert and Gibbons-Hawking term, and the third term is Mann and Marolf counterterm (MM-counterterm, henceforth) defined from (\ref{MMrltn}). The fact that the on-shell action (\ref{ACTwtCC}) is finite and the variation of the action vanishes on-shell was proved in {\cite{Mann:2006}} for both cases of cylindrical cut-off and hyperbolic cut-off. In this section, we compute the variation of the action (\ref{ACTwtCC}) with respect to $h_{ab}$ and the form of the boundary stress tensor.

Taking a variation of the action with respect to the metric and eliminating the equation of the motion, we get
\begin{equation}
\delta S = \frac{1}{16 \pi G} \int_{\partial {\cal{M}}} \sqrt{-h} \bigg( - \pi^{ab} + \hat{\pi}^{ab} + \Delta^{ab} \bigg) \delta h_{ab} \label{vrtnAC}
\end{equation}
where $\pi^{ab} = K^{ab} - K h^{ab}$, $\hat{\pi}^{ab} = \hat{K}^{ab} - \hat{K} h^{ab}$, and $\Delta^{ab}$ represents extra terms
that arise from the definition   (\ref{MMrltn}). Explicitly \cite{Mann:2008}
\begin{equation}
\Delta^{ab} = \hat{K}^{ab} - 2 \tilde{L}^{cd} \bigg( \hat{K}_{cd} \hat{K}^{ab} - \hat{K}^{a}_{c} \hat{K}^{b}_{d} \bigg)
+ {\cal{D}}^2 \tilde{L}^{ab} + h^{ab} {\cal{D}}_{k} {\cal{D}}_{l} \tilde{L}^{kl} - {\cal{D}}_{k} \bigg( {\cal{D}}^{a} \tilde{L}^{kb} + {\cal{D}}^{b} \tilde{L}^{ka} \bigg) \label{dlt_ab}
\end{equation}
here ${\cal{D}}_{a}$ is a covariant derivative compatible with $h_{ab}$ defined on $n+2$ dimensional hypersurface, and $\tilde{L}^{ab}$ indicates
\begin{eqnarray}
\label{rgnL} L^{\; \; \; ab}_{cd} &=& h^{ab} \hat{K}_{cd} + \delta^{a}_{c} \delta^{b}_{d} \hat{K} - \delta^{a}_{c} \hat{K}^{b}_{d} - \delta^{b}_{c} \hat{K}^{a}_{d},  \\
\label{tldL} \tilde{L}^{ab} &=& h^{cd} (L^{-1})^{\; \; \; ab}_{cd}.
\end{eqnarray}
The detailed procedures are described in appendices (A) and (D). Equation (\ref{vrtnAC}) directly leads to the boundary stress tensor, which is defined as the functional derivative of the on-shell action with respect to $h_{ab}$, associated with MM-counterterm, and is
\begin{equation}
T_{ab} := - \frac{2}{\sqrt{-h}} \frac{\delta S}{\delta h^{ab}} = \frac{1}{8 \pi G} \bigg( \pi_{ab} - \hat{\pi}_{ab} + \Delta_{ab} \bigg) \label{BStnsr}
\end{equation}
where only $T^{\pi}_{ab}$, which indicates the first two terms in the right side, is expected to yield conserved charges. The explicit form of $\Delta^{ab}$ will be obtained in section 3.

%We define a part of the boundary stress tensor contributed by $\Delta_{ab}$ as $T^{\Delta}_{ab}$

%----------------------------------------------
\subsection{Asymptotic Flatness}
\label{sec:AsmyFlt}
%----------------------------------------------

Adopting an approach to defining asymptotic flatness similar to that in hyperbolic coordinates  {\cite{Beig:1982}} {\cite{Beig:1984}}, we define a spacetime $(\mathcal{M}, g)$ in cylindrical coordinates and confine ourselves to this spacetime throughout this paper. Assuming that a static spacetime $(\mathcal{M}, g)$ is radially smooth of order $m$ at spatial infinity in $(n+3)$ dimensions, the components of the metric take the asymptotic form
\begin{equation}\label{Orgmtr}
g_{\mu \nu} = \eta_{\mu \nu} + \sum^{m}_{k=1} \frac{l^{(k)}_{\mu \nu}(\eta^A/r)}{r^{n+k-1}} + f^{(m+1)}_{\mu \nu}(r,\eta^A)
\end{equation}
where $n \geq 1$,  $r$ is a radial coordinate, and $\eta^A$ are angular coordinates associated with the metric $\mu^{(0)}_{AB}$ on the unit sphere $S^{n+1}$, and $l^{(k)}_{\mu \nu}$ is $C^{\infty}$ in $\eta^{A}/r$ and $f^{(k)} = {\mathcal{O}}(1/r^m)$. Defining functions $w^{a}(\eta^A)$ at $t=const.$ such that
\begin{equation}
\frac{x^a}{r} = w^a(\eta^A), \; \; \; \; \; dx^a = w^a dr + r w^a_{,A} d \eta^{A},
\end{equation}
(\ref{Orgmtr}) transforms into
\begin{align}
& \eta_{\mu \nu} dx^{\mu} dx^{\nu} = - dt^2 + dr^2 + r^2 \mu^{(0)}_{AB} d \eta^{A} d \eta^{B}, \nonumber\\
& \tilde{\gamma}^{(k)} = - l_{tt}^{(k)}, \; \; \; \; \tilde{\alpha}^{(k)} = l^{(k)}_{ab} w^a w^b, \; \; \; \; J^{(k)}_{A} = l^{(k)}_{ab} w^a w^b_{,A}, \; \; \; \; \mu^{(k)}_{AB} = l^{(k)}_{ab} w^{a}_{,A} w^{b}_{,B}
\end{align}
in turn yielding the explicit form
\begin{align} \label{Trnfmtr}
ds^2 =& - \bigg( 1 + \sum^{m}_{k=1} \frac{\tilde{\gamma}^{(k)}(\eta^{A})}{r^{n+k-1}} + {\mathcal{O}} \bigg(\frac{1}{r^{m+1}} \bigg) \bigg)dt^2 + \bigg( 1 + \sum^{m}_{k=1} \frac{\tilde{\alpha}^{(k)}(\eta^{A})}{r^{n+k-1}} + {\mathcal{O}} \bigg(\frac{1}{r^{m+1}} \bigg) \bigg)dr^2 \nonumber\\
&+ 2 \bigg( \sum^{m}_{k=1} \frac{J^{(k)}_{A} (\eta^{A})}{r^{n+k-1}} + {\mathcal{O}} \bigg(\frac{1}{r^{m+1}} \bigg) \bigg)r dr d \eta^A + r^2 \bigg(\mu^{(0)}_{AB} + \sum^{m}_{k=1} \frac{ \mu^{(k)}_{AB}(\eta^{A})}{r^{n+k-1}} + {\mathcal{O}} \bigg(\frac{1}{r^{m+1}} \bigg) \bigg) d \eta^{A} d \eta^{B}
\end{align}
where ${\tilde{\gamma}}^{(k)}$, ${\tilde{\alpha}}^{(k)}$ are smooth functions, and $J^{(k)}_{A}$ are smooth vector fields, and $\mu^{(1)}_{AB}$, $\mu^{(2)}_{AB}$ are smooth tensor fields on $S^{n+1}$. The symbols ${\mathcal{O}}(r^{-(m+1)})$ refer to terms that fall-off at least as fast as $r^{-(m+1)}$ as one approaches spacelike infinity, i.e., $r \rightarrow + \infty$ with fixed $\eta$. Without loss of generality, we find it convenient to substitute
\begin{equation}
\bigg(1 + \sum^{m}_{k=1} \frac{\tilde{\alpha}^{(k)}}{r^{n+k-1}} \bigg) = \bigg(1 + \sum^{m}_{k=1} \frac{\alpha^{(k)}}{r^{n+k-1}} \bigg)^2 + {\mathcal{O}} \bigg(\frac{1}{r^{m+1}} \bigg)
\end{equation}
in (\ref{Trnfmtr}), and likewise for the $g_{tt}$-component ( $\tilde{\gamma}$ changes to $\gamma$ ). In order to simplify the metric, we first try to remove $J^{(1)}_{A}$ in (\ref{Trnfmtr}) by using a coordinate transformation
\begin{align} \label{cdntrf1}
& \eta^{A} = \bar{\eta}^{A} + \frac{1}{r^n} G^{(1)A}(\bar{\eta}^{B}), \; \; \; \; \; r=\bar{r}, \; \; \; \; \; t=t, \nonumber\\
& d \eta^{A} = d \bar{\eta}^{A} + \frac{1}{r^n} G^{(1)A}_{,B} d \bar{\eta}^{B} -  \frac{n}{r^{n+1}} G^{(1)A} dr.
\end{align}
Applying (\ref{cdntrf1}) into (\ref{Trnfmtr}), the leading term of the $g_{rA}$-component is eliminated by choosing
\begin{equation}
J^{(1)}_{A} = n G^{(1)B} \mu^{(0)}_{AB},
\end{equation}
and this allows us to set $J^{(1)}_{A} = 0$ in (\ref{Trnfmtr}). Subsequently we get rid of $\alpha^{(2)}$ via the additional coordinate transformation :
\begin{align}\label{cdntrf1a}
& r = \bar{r} + \frac{1}{\bar{r}^n} F^{(2)}(\eta^A), \nonumber\\
& dr = d\bar{r} - n \frac{1}{\bar{r}^{n+1}} F^{(2)} d \bar{r} + \frac{1}{\bar{r}^n} F^{(2)}_{,B} d \eta^B.
\end{align}
Plugging these to (\ref{Trnfmtr}), the $1/\bar{r}^{n+1}$-term in $d\bar{r}^2$ can be set to zero via
\begin{equation}
\alpha^{(2)} = n F^{(2)}
\end{equation}
where the leading term in $r dr d \eta^{A}$ is not affected. Generalizing these coordinate transformations to include higher orders of $1/r$ yields
\begin{align}
& \eta^{A} = \bar{\eta}^{A} + \frac{1}{\bar{r}^{n+k-1}} G^{(k)A}, \; \; \; \; r = \bar{r} + \frac{1}{\bar{r}^{n+k-1}} F^{(k+1)}
\end{align}
where these transformations are sequentially applied to the above to the metric.  We can then show that
\begin{align}
&J^{(1)}_{A} = J^{(2)}_{A} = \cdots = J^{(m)}_{A} = 0, \nonumber\\
&\alpha^{(2)} = \alpha^{(3)} = \cdots = \alpha^{(m)} = 0.
\end{align}
We finally obtain the simplified form of the metric
\begin{align}
ds^2=& \bigg( 1 + \frac{\alpha}{r^n} \bigg)^2 dr^2 - \bigg( 1 + \frac{\gamma^{(1)}}{r^n} + \frac{\gamma^{(2)}}{r^{n+1}} + \mathcal{O} \bigg( \frac{1}{r^{n+2}} \bigg) \bigg)^{2} dt^2 \nonumber\\
&+ r^2 \bigg( \mu^{(0)}_{AB} + \frac{1}{r^n} \mu^{(1)}_{AB} + \frac{1}{r^{n+1}} \mu^{(2)}_{AB} + \mathcal{O} \bigg( \frac{1}{r^{n+2}} \bigg)  \bigg) d \eta^{A} d \eta^{B} \nonumber,
\end{align}
which we rewrite as
\begin{align}
ds^2 =& N^2 dr^2 + h_{ab} dx^{a} dx^{b} \nonumber\\
=& \bigg( 1 + \frac{\alpha}{r^n} \bigg)^2 dr^2 + \bigg( h^{(0)}_{ab} + \frac{1}{r^n} h^{(1)}_{ab} + \frac{1}{r^{n+1}} h^{(2)}_{ab} + \cdots \bigg) dx^{a} dx^{b} \label{Mtr}
\end{align}
where $x^a = (t,\eta^A)$ are coordinates on the $(n+2)$-dimensional hypersurface compatible with the metric $h_{ab}$, whose expansion is
\begin{align}
h_{ab} dx^a dx^b = - \bigg( 1 + \frac{\gamma^{(1)}}{r^n} + \frac{\gamma^{(2)}}{r^{n+1}} + \mathcal{O} \bigg( \frac{1}{r^{n+2}} \bigg) \bigg)^{2} dt^2 + r^2 \bigg( \mu^{(0)}_{AB} + \frac{1}{r^n} \mu^{(1)}_{AB} + \frac{1}{r^{n+1}} \mu^{(2)}_{AB} + \mathcal{O} \bigg( \frac{1}{r^{n+2}} \bigg)  \bigg) d \eta^{A} d \eta^{B}
\end{align}
where $a=t, A$.

The boundary spacetime $(\partial{\mathcal{M}},h)$ is actually a one-parameter family $(\mathcal{M}_{\Omega}, g_{\Omega})$ where $\mathcal{M}_{\Omega} \subset \mathcal{M}$ and  $\mathcal{M}_{\Omega}$ converges to $\mathcal{M}$ with increasing $\Omega$. The boundary of a region $(\mathcal{M}_{\Omega}, g_{\Omega})$ for a certain value of $\Omega$ is described by $(\partial \mathcal{M}_{\Omega}, h_{\Omega})$. As $\Omega$ is varied, we get a family of boundaries that
provide a specific way of `cutting-off' the space-time  ${\mathcal{M}}$, with the asymptotic boundary obtained as $\Omega \rightarrow \infty$. In this paper, our interest is in the class  ``cylindrical cut-offs", for which
\begin{equation}
\Omega^{cyl} = r + O(r^{0}).
\end{equation}
Note that the metric (\ref{Mtr}) takes the same form as the metric in  hyperbolic coordinates {\cite{Mann:2008}} except that $h_{ab}$ is further decomposed  into a $tt$-component and angular components. Naively  one might expect that our result is easily derived from the hyperbolic case in ref. {\cite{Mann:2008}} where $\Omega$ is taken to be
\begin{equation}
\Omega^{hyp} = \rho + O(\rho^{0}).
\end{equation}
and the coordinate $\rho$ is defined by $\rho^2 = r^2 - t^2$. However this is not true since in hyperbolic coordinates the boundary metric
(parametrized by a surface $\rho$=constant)  is manifestly covariant under a variation, whereas in cylindrical coordinates the boundary metric defined at $r=$constant does not change fully covariantly.  In particular, the $tt$-component and angular components in the induced metric $h_{ab}$ are have expansions in  different orders of $r$. For example, taking a variation with respect to each surface parameter, i.e. $\rho$ or $r$, in hyperbolic coordinates
\begin{equation}
\frac{\partial}{\partial \rho} \bigg(\rho h^{(0)}_{ab} \bigg) = h^{(0)}_{ab}, \label{vrthycrd}
\end{equation}
whereas in   cylindrical coordinates
\begin{equation}
\frac{\partial}{\partial r} \bigg( r h^{(0)}_{ab} \bigg) = h^{(0)}_{ab} + 2 r^2 \mu^{(0)}_{ab}. \label{vrtclnd}
\end{equation}
In (\ref{vrtclnd}) only the angular components transform covariantly in the $(n+1)$ dimensional subspace, whereas in
 (\ref{vrthycrd}), all components transform covariantly.  This distinction introduces new features and subtleties in cylindrical coordinate that are rather more complicated than  the hyperbolic case.

%----------------------------------------------
%----------------------------------------------
\section{Boundary Stress Tensor $T^{ab}$}
\label{sec:CalDelta}
%----------------------------------------------
%----------------------------------------------

With the definition of asymptotic flatness in (\ref{Mtr}), we investigate $\Delta^{ab}$ by finding its definite form. Note that $\Delta^{ab}$ is a remnant term, obtained from subtracting $\hat{\pi}^{ab}$ from the variation of the MM-counterterm in the action, and so is totally expressed by the MM-counterterm solution $\hat{K}^{ab}$ and its covariant derivatives. Consequently, we shall see that $\Delta^{ab}$ does not vanish entirely.
%but take a form except $[\Delta^{ab}]^{(1)}$ in $n=1$ case. \tcb{\bf [I don't understand that last sentence.]}
We next consider the role that the non-vanishing $\Delta^{ab}$ play with respect to conserved quantities.

\subsection{Calculation of $\Delta^{ab}$}

We start with setting up the spacelike normal vector $n^{\alpha}$ on a cylindrical hypersurface on $r=$constant, where asymptotically $r \rightarrow \infty$, and calculate its extrinsic curvature $K_{ab}$. At the boundary  the decomposed Einstein equations are  \begin{eqnarray}
\label{DCEE1} \perp(R_{ab}) &=& {\cal{R}}_{ab} + {\cal{D}}_{a} a_{b} - a_{a}a_{b} - \pounds_{n}K_{ab} - KK_{ab} + 2 K_{a}^{c}K_{cb}, \\
\label{DCEE2} \perp ({R_{ac}n^{c}}) &=& {\cal{D}}^{b} K_{ab} - {\cal{D}}_{a} K = - {\cal{D}}^{b} \pi_{ab}, \\
\label{DCEE3} R_{ab} n^{a} n^{b} &=& -\pounds_{n} K - K^{ab} K_{ab} + ({\cal{D}}_{b}a^{b} - a^{b}a_{b}),
\end{eqnarray}
where $a^{b}$ and $K_{ab}$ are defined by
\begin{equation}
a^{b} = n^{a} \nabla_{a} n^{b}, \; \; \; \; \; K_{ab} = \nabla_{a} n_{b} - n_{a} a_{b},
\end{equation}
and the last equation  can be rewritten  as
\begin{equation}
{\cal{R}} - K^2 + K^{ab}K_{ab} = 0.
\end{equation}
For  asymptotically flat spacetimes, the left-hand sides of  (\ref{DCEE1}) $-$ (\ref{DCEE3}) become zero as $r \rightarrow \infty$.  Equation  (\ref{DCEE1}) yields ${\mathcal{R}}_{ab}$. The remaining equations yield constraint conditions between the coefficients in the metric, e.g. $\alpha$, $\gamma^{(1)}$ or $\gamma^{(2)}$. We solve the decomposed Einstein equation in powers of $1/r$, i.e.
$\mathcal{R}_{ab} = \mathcal{R}^{(0)}_{ab} + \frac{1}{r^n} \mathcal{R}^{(1)}_{ab} + \frac{1}{r^{n+1}} \mathcal{R}^{(2)}_{ab} + \cdots$, where for $n=1$ the sub-sub-leading term must be separately dealt with from the $n \geq 2$ cases. This is because when $n=1$ the product of two sub-leading  terms  has the same order as the sub-sub-leading term (both of order of $1/r^2$), whereas for $n \geq 2$ this product is of order of $1/r^{2n}$ and falls off faster than the  sub-sub-leading term (of order $1/r^{n+1}$) and so does not contribute  at that  order. The solutions and process are described in appendix B.

Now, for convenience we redefine $\hat{K}_{ab}$ in terms of $\hat{Q}_{ab}$
\begin{equation}
\hat{K}_{ab} = r \hat{Q}_{ab} = r \hat{Q}^{(0)}_{ab} + \frac{1}{r^{n-1}} \hat{Q}^{(1)}_{ab} + \frac{1}{r^n} \hat{Q}^{(2)}_{ab} + \cdots, \label{MMcntQ}
\end{equation}
and then expand relation (\ref{MMrltn}):
\begin{align}
        \label{MMrltnep1} {\mathcal{R}}^{(0)}_{ab} =& n \mu^{(0)}_{ab}, \\
        \label{MMrltnep2} {\mathcal{R}}^{(1)}_{ab} =& r^2 \hat{Q}^{(1)} \mu^{(0)}_{ab} + (n-1) \hat{Q}^{(1)}_{ab} - 2 u^{(0)}_{\; \; a} u^{(0)c} \hat{Q}^{(1)}_{cb} - \mu^{(1)} \mu^{(0)}_{ab} + \mu^{(1)}_{ab}, \\
        \label{MMrltnep3} (n \geq 2), \; \; {\mathcal{R}}^{(2)}_{ab} =& r^2 \hat{Q}^{(2)} \mu^{(0)}_{ab} + (n-1) \hat{Q}^{(2)}_{ab} - 2 u^{(0)}_{\; \; a} u^{(0)c} \hat{Q}^{(2)}_{cb} - \mu^{(2)} \mu^{(0)}_{ab} + \mu^{(2)}_{ab},\\
        \label{MMrltnep4} (n=1), \; \; {\mathcal{R}}^{(2)}_{ab} =& r^2 \hat{Q}^{(2)} \mu^{(0)}_{ab} - \mu^{(2)} \mu^{(0)}_{ab} + \mu^{(2)}_{ab} - 2 u^{(0)}_{\; \; a} u^{(0)c} \hat{Q}^{(2)}_{\; bc} + \frac{r^2}{4} \mathcal{R}^{(1)} \mathcal{R}^{(1)}_{ab} + \frac{r^2}{4} {\mathcal{R}}^{(1)} \mu^{(1)}_{ab} \nonumber\\
        &- \frac{1}{4} \mu^{(1)} {\mathcal{R}}^{(1)}_{ab} - \frac{r^2}{2} {\mathcal{R}}^{(1)}_{cd}h^{(1)cd} \mu^{(0)}_{ab} - \frac{r^2}{4} {\mathcal{R}}^{(1)c}_{\; \; a}{\mathcal{R}}^{(1)}_{cb} +  \frac{1}{4} {\mathcal{R}}^{(1)}_{ae} \mu^{(1)e}_{\; \; b} + \frac{1}{4} \mu^{(1)e}_{\; \; a} {\mathcal{R}}^{(1)}_{eb} \nonumber\\
        &- \frac{1}{4} \mu^{(1)} \mu^{(1)}_{ab} + \frac{1}{2} \mu^{(1)}_{cd} \mu^{(1)cd} \mu^{(0)}_{ab} - \frac{1}{4} \mu^{(1)c}_{\; \; a} \mu^{(1)}_{cb}
\end{align}
where the sub-sub-leading order is separately treated as noted above and $\hat{Q}^{(i)}$ is the trace of $\hat{Q}^{(i)}_{ab}$ . After some rearrangement  and insertion of the asymptotic expansion of the metric (with details given in appendix C) we obtain $\hat{Q}_{ab}$ explicitly  for $n \geq 2$:
\begin{align}
\hat{Q}^{(0)}_{ab} = & \mu^{(0)}_{ab}, \\
\label{qab1frn} \hat{Q}^{(1)}_{ab} = &  \frac{1}{(n-1)}\bigg[ (n-1)\mu^{(1)}_{ab} + \alpha \mu^{(0)}_{ab} + \gamma^{(1)} \mu^{(0)}_{ab} + D_{a}D_{b} \alpha \bigg], \\
\label{qab2frn} \hat{Q}^{(2)}_{ab} = & \frac{1}{(n-1)} \bigg[ \frac{(n-1)}{2r^2} h^{(2)}_{ab} + \frac{(n+2)}{n} \gamma^{(2)} \mu^{(0)}_{ab} + n \mu^{(2)}_{ab}\bigg].
\end{align}
For $n=1$ it is clear from equations (\ref{MMrltnep2}) and (\ref{MMrltnep4}) that  $\hat{Q}^{(i)}_{ab}$ for $i=1,2$ cannot be uniquely determined.  As shown in appendix C, we can determine  $\hat{Q}^{(i)}_{ab}$ up certain ambiguities; explicitly we find
$$
\hat{Q}^{(1)}_{ab} =
\beta_1 \mathcal{R}^{(1)}_{ab} + r^2 \beta_2 \mathcal{R}^{(1)} \mu^{(0)}_{ab} + \lambda_{1} \mu^{(1)}_{ab} + \lambda_2 \mu^{(1)} \mu^{(0)}_{ab}
$$
where $\beta_1 + 2 \beta_2 = \frac{1}{2}$, $\lambda_1 + 2 \lambda_2 = \frac{1}{2}$.   The ambiguities can be eliminated by applying
$\mu^{(1)}_{ab} = - 2 \alpha \mu^{(0)}_{ab}$, which is the condition for fixing asymptotic supertranslation symmetry, and using $D_{a} D_{b} \alpha = - \alpha \mu^{(0)}_{ab} - \gamma^{(1)} \mu^{(0)}_{ab}$ as shown in appendix C. As a result, we obtain
\begin{equation}
\label{qab1fr1} \hat{Q}^{(1)}_{ab} = \frac{1}{2} \mu^{(1)}_{ab} + D_{a} D_{b} \alpha
\end{equation}

The expression for $\hat{Q}^{(2)}_{ab}$ is somewhat more complicated (see eq. (\ref{antzQ12})) and is determined up to terms of the form $\lambda_{3} \mu^{(2)}_{ab} + \lambda_4 \mu^{(2)} \mu^{(0)}_{ab}$ where $\lambda_3 + 2 \lambda_4 = \frac{1}{2}$.  We can choose the remaining coefficients by demanding that the relation (\ref{cnsvdcg}) holds; as shown in appendix C, this gives $\lambda_3= -\frac{3}{2}$ and $\lambda_4 = \frac{1}{2}$, thereby yielding
\begin{equation}
\label{qab2fr1} \hat{Q}^{(2)}_{ab} = - \frac{1}{2} \mu^{(2)}_{ab} + \frac{1}{2} \mu^{(2)} \mu^{(0)}_{ab} - \frac{3}{2} \gamma^{(2)} \mu^{(0)ab} - \frac{1}{r^2} \gamma^{(2)} u^{(0)}_{a} u^{(0)}_{b} - \frac{5}{2} \alpha^2 \mu^{(0)}_{ab} + \frac{2}{r^2} \alpha \gamma^{(1)} u^{(0)a} u^{(0)b}.
\end{equation}

Next, the form of $\tilde{L}^{ab}$ in (\ref{dlt_ab}) can be found by using (\ref{rgnL}) and (\ref{tldL}); we find for $n \geq 2$
\begin{align}
\tilde{L}^{(0)ab} =& \frac{r}{n(n+1)}\bigg[ \frac{(n+1)}{2r^2} \mu^{(0)ab} - \frac{(n-1)}{2} u^{(0)a} u^{(0)b} \bigg], \\
\label{tldL1frn} \tilde{L}^{(1)ab} =& \frac{r}{n(n-1)}\bigg[ \frac{r^2}{(n-1)} D^a D^b \alpha  { - \frac{(n-1)}{2 r^2} \mu^{(1)ab}} + \frac{(n^2+1)}{2(n-1)r^2} \alpha \mu^{(0)ab} + \frac{(n^2+1)}{2(n-1)r^2} \gamma^{(1)} \mu^{(0)ab} \nonumber\\
& + \frac{n}{4r^2} \mu^{(1)} \mu^{(0)}_{ab} - \frac{(n-1)^2}{2(n+1)} \alpha u^{(0)a} u^{(0)b} + \frac{(n-1)^2}{2(n+1)} \gamma^{(1)} u^{(0)a} u^{(0)b} - \frac{n(n-1)^2}{4(n+1)^2} \mu^{(1)} u^{(0)a} u^{(0)b} \bigg], \\
\label{tldL2frn} \tilde{L}^{(2)ab} =& \frac{r}{n(n-1)} \bigg[\frac{(2n^2-5n+3)}{2(n+1)} \gamma^{(2)} u^{(0)a} u^{(0)b}  { -\frac{(n^3-4n^2-5n-4)}{2n(n-1)(n+1)r^2} \gamma^{(2)} \mu^{(0)ab}}-\frac{n(n-3)}{2(n-1)r^2} \mu^{(2)ab} \bigg],
\end{align}
and for $n=1$
\begin{align}
\tilde{L}^{(0)ab} &= \frac{1}{2r} \mu^{(0)ab}, \; \; \; \; \; \; \; \; \; \label{tldL1fr1} \tilde{L}^{(1)ab} = \frac{1}{r} \alpha \mu^{(0)ab} + \frac{1}{2r} \gamma^{(1)} \mu^{(0)ab}, \\
\label{tldL2fr1} \tilde{L}^{(2)ab} &= {-\frac{1}{4r} \mu^{(2)ab}} + \frac{1}{2r} \gamma^{(2)} \mu^{(0)ab} - \frac{r}{4} \gamma^{(2)} u^{(0)a} u^{(0)b} + \frac{13}{4r} \alpha^2 \mu^{(0)ab} \nonumber\\
& \; \; \; + \frac{5}{2r} \alpha \gamma^{(1)} \mu^{(0)ab} + \frac{1}{2r} (\gamma^{(1)})^2 \mu^{(0)ab} + \frac{r}{2} \alpha \gamma^{(1)} u^{(0)a} u^{(0)b}.
\end{align}
This process is described in appendix D.

Recalling from (\ref{dlt_ab}) the form of $\Delta^{ab}$
\begin{displaymath}
\Delta^{ab} = \hat{K}^{ab} - 2 \tilde{L}^{cd} \bigg( \hat{K}_{cd} \hat{K}^{ab} - \hat{K}^{a}_{c} \hat{K}^{b}_{d} \bigg)
+ {\cal{D}}^2 \tilde{L}^{ab} + h^{ab} {\cal{D}}_{k} {\cal{D}}_{l} \tilde{L}^{kl} - {\cal{D}}_{k} \bigg( {\cal{D}}^{a} \tilde{L}^{kb} + {\cal{D}}^{b} \tilde{L}^{ka} \bigg),
\end{displaymath}
and expanding
\begin{align}
    {\Delta}^{ab} =[{\Delta}^{ab}]^{(0)} + \frac{1}{r^n} [{\Delta}^{ab}]^{(1)} + \frac{1}{r^{n+1}} [{\Delta}^{ab}]^{(2)} + \cdots,
\end{align}
we find for $ n>1 $  the leading order and the  sub-leading orders to be
\begin{align}
\label{Dltn0} [\Delta^{ab}]^{(0)}  =& 0, \\
\label{Dltn1} [\Delta^{ab}]^{(1)}  =& \frac{r}{n(n-1)} \bigg[D^a D^b \gamma^{(1)} + \frac{n^2}{(n+1)^2} D^2 \mu^{(1)} u^{(0)a} u^{(0)b} + \frac{2n}{(n+1)} D^2 \alpha u^{(0)a} u^{(0)b}  \bigg]
\end{align}
Indeed, that $[\Delta^{ab}]^{(1)}$ is non-zero is quite obvious, because from (\ref{dlt_ab})
the indices of the first three terms in $\Delta^{ab}$ contain only angular components (note $\hat{K}^{(1)}_{ab} = r \hat{Q}^{(1)}_{ab}$), whereas the
 last four terms  (beginning with $\tilde{L}^{(1)}_{ab}$ in (\ref{dlt_ab}) ) have both angular components and  $tt$-components. As a result, in the summation, the angular parts are canceled out except $D^a D^b \gamma^{(1)}$, but $tt$-component terms remain.

 At sub-sub-leading order we obtain
\begin{align}
\label{Dltn2} [\Delta^{ab}]^{(2)} = \frac{r}{n(n-1)^2} \bigg[\frac{(n^3-2n^2-n-2)}{n(n+1)} D^a D^b \gamma^{(2)} - \frac{(n^3+8n^2+5n+2)}{n(n+1)r^4} \gamma^{(2)}\mu^{(0)ab} - \frac{n(n+1)}{r^4} \mu^{(2)ab} \bigg].
\end{align}
In contrast to the sub-leading case,  in the sub-sub-leading term both   $\hat{K}^{(2)}_{ab}$ ($= r \hat{Q}^{(2)}_{ab}$, given in (\ref{qab2frn})) and $\tilde{L}^{(2)}_{ab}$ (given in (\ref{tldL2frn})) carry both $tt$ and angular components.  Here we see similarities with the hyperbolic case  {\cite{Mann:2008}}, for which $\hat{K}^{(2)}_{ab} = \rho h^{(2)}_{ab}$ and $h^{(2)}=0$ were respectably obtained  from the MM-relation and the decomposed Einstein equations. These require $\mathcal{R}^{(2)} = 0$, and  yield $[\Delta^{ab}]^{(2)}=0$.   For the cylindrical case we are considering, we find that
the terms associated with $h^{(2)}_{ab}$ in $\hat{K}^{(2)}_{ab}$ and in $\tilde{L}^{(2)}_{ab}$ cancel out in $[\Delta^{ab}]^{(2)}$, as shown in appendix E. However, unlike the hyperbolic case, in the cylindrical case, $h^{(2)}$ breaks up into $\mu^{(2)}$ and $\gamma^{(2)}$ (explicitly $h^{(2)} = \mu^{(2)} + 2 \gamma^{(2)}$); consequently the decomposed  Einstein equations imply $\mathcal{R}^{(2)}$ is non-vanishing and contributes to
$\hat{K}^{(2)}_{ab}$.  Indeed from eq. (\ref{CNSTcdt2forn}) we see that $\mathcal{R}^{(2)}$ can be expressed in terms of any one of
$h^{(2)}$, $\mu^{(2)}$, or $\gamma^{(2)}$; we have expressed the result in terms of $\gamma^{(2)}$ in (\ref{qab2frn}).
 The quantity $\gamma^{(2)}$ does not  vanish, but remains in (\ref{Dltn2}).

For $n=1$ up to  sub-leading order we find
\begin{align}
& [\Delta^{ab}]^{(0)} = 0, \; \; \; \; \; \textrm{and} \; \; \; \; \; \label{Dlt11} [\Delta^{ab}]^{(1)} = 0,
\end{align}
where we have used $D_{a} D_{b} \alpha = - \alpha \mu^{(0)}_{ab} - \gamma^{(1)} \mu^{(0)}_{ab}$, which can be  inferred from $D^2 \alpha = - \frac{2}{r^2} \alpha - \frac{2}{r^2} \gamma^{(1)} $. The sub-sub-leading term is
\begin{align}
\label{Dlt12} & [\Delta^{ab}]^{(2)} = - \frac{2}{r^3} \mu^{(2)ab} - \frac{2}{r^3} \gamma^{(2)} \mu^{(0)ab} - \frac{4}{r^3} \alpha^2 \mu^{(0)ab} - \frac{6}{r^3} \alpha \gamma^{(1)} \mu^{(0)ab} - \frac{4}{r^3} (\gamma^{(1)})^2 \mu^{(0)ab} \nonumber\\
& + \frac{9}{2r} D^e \alpha D_{e} \alpha \mu^{(0)ab} + r D^a D^b \gamma^{(2)} - 5 r D^a \alpha D^b \alpha  - \frac{3r}{2} D^2 \alpha^2 u^{(0)a} u^{(0)a} - \frac{5r}{4} D^2 (\alpha \gamma^{(1)}) u^{(0)a} u^{(0)b}.
\end{align}
These results in $n=1$ are commensurate with the hyperbolic case \cite{Mann:2008}, which has manifestly vanishing $[\Delta^{ab}]^{(0)}$ and $[\Delta^{ab}]^{(1)}$, but non-vanishing $[\Delta^{ab}]^{(2)}$.

That we find  $\Delta^{ab}$ non-vanishing implies that the boundary stress tensor in cylindrical coordinates generally takes  the form $T_{ab}$ in (\ref{BStnsr}) and not $T^{\pi}_{ab}$ in (\ref{BStnsr_pi}).

\subsection{Conserved Quantities and $\Delta^{ab}$}

Since the boundary stress tensor is described not by $T^{\pi}_{ab}$ but by $T_{ab}$  (due  $\Delta^{ab}\neq 0$),  we now consider how $\Delta^{ab}$ is related to  conserved quantities as given in equation (\ref{cnsvdcg}). Plugging $T_{ab} = T^{\pi}_{ab} - \Delta_{ab}$ into (\ref{cnsvdcg}), we see that $\Delta^{ab}$ will contribute to  conserved quantities  via
\begin{equation}
Q^{\Delta}[\xi] = - \frac{1}{8 \pi G} \oint d^{n+1} x \sqrt{\gamma} u^{a} \Delta_{ab} \xi^{b}.
\end{equation}

For  $n \geq 2$  we find that $Q^{\Delta}[\xi] = 0$.  The sub-leading term contributes
\begin{align}\label{Q1}
    [Q^{\Delta}]^{(1)} &= - \frac{1}{8 \pi G} \oint d^{n+1} x \sqrt{\gamma} u^{(0)a} [\Delta_{ab}]^{(1)} \xi^{(0)b} , \nonumber\\
                       &= \frac{r}{8 \pi G} \frac{1}{(n-1)(n+1)} \oint d^{n+1} x \sqrt{\gamma} \bigg(\frac{n}{(n+1)} D^2 \mu^{(1)} + 2 D^2 \alpha \bigg), \nonumber\\
                       &= 0,
\end{align}
where $u^{(0)a} = - \delta^{a}_{t}$ and $\xi^{(0)t} = 1$ has been used, and the total derivative on the closed surface becomes zero.  The sub-sub-leading order also makes no contribution, since from
(\ref{Dltn2}) we see that $[\Delta_{ab}]^{(2)}$ contracted with the timelike normal vector $u^{(0)a}$ vanishes and so $[Q^{\Delta}]^{(2)}$ obviously becomes zero.

For $n=1$  $[\Delta^{ab}]^{(1)}=0$ and so only the sub-sub-leading term could possibly contribute.  Carrying out similar manipulations to the previous case, we get
\begin{align}\label{Q2}
    [Q^{\Delta}]^{(2)} &= - \frac{1}{8 \pi G} \oint d^{2} x \sqrt{\gamma} u^{(0)a} [\Delta_{ab}]^{(2)} \xi^{(0)b} , \nonumber\\
                       &= - \frac{1}{8 \pi G} \oint d^{2} x \sqrt{\gamma} \bigg( \frac{3r}{2} D^2 \alpha^2 + \frac{5r}{4} D^2 (\alpha \gamma^{(1)}) \bigg), \nonumber\\
                       &= 0,
\end{align}
where $u^{(0)a} = - \delta^{a}_{t}$ and $\xi^{(0)t} = 1$ has been used, and the total derivative on the closed surface becomes zero at the end. Indeed, this result is expected, because we required that {$\hat{Q}_{ab}^{(2)}$} not  contribute to conserved charges.

Hence $Q^{\Delta} = 0$ even though $\Delta^{ab} \neq 0$. As a result the conserved quantity formula is of the form  (\ref{cnsvdcg}) and is given only in terms of $T^{\pi}_{ab}$.

%----------------------------------------------
%----------------------------------------------
\section{$(n+3)$-dimensional Static Spacetime}
\label{sec:Cnschg}
%----------------------------------------------
%----------------------------------------------

In this section, we apply the boundary stress tensor method to $(n+3)$-dimensional static spacetime. We show that $\Delta^{ab}$ makes no contribution with respect to the conserved quantities and obtain the conserved charges by using (\ref{cnsvdcg}).

We examine the boundary stress tensor method associated with the MM-counterterm in $(n+3)$-dimensional static spacetime. In this spacetime, we check that $\Delta^{ab}=0$, and prove that the boundary stress tensor yields conserved charges agreed with the usual definition {\cite{Arnowitt:1960}} {\cite{Myers:1986}}.

From the Myers-Perry static black hole solution {\cite{Myers:1986}}, the metric is
\begin{align}
ds^2 = - \bigg(1 - \frac{\mu}{r^{n+2}} \bigg) dt^2 + \bigg( 1- \frac{\mu}{r^{n+2}} \bigg)^{-1} dr^2 + r^2 d \Omega^2_{n+1}, \label{Mpstt}
\end{align}
where $\mu$ is related to the mass $M$
\begin{equation}
M = \frac{(n+2) A_{n+1}}{16 \pi G} \mu, \; \; \; \; \; \; A_{n+1} = \frac{2 \pi^{(n+2)/2}}{\Gamma((n+2)/2)},
\end{equation}
and comparing (\ref{Mpstt}) with our metric (\ref{Mtr}), they are related to
\begin{equation}
\gamma^{(1)} = -\frac{1}{2} \mu, \; \; \; \; \; \alpha = \frac{1}{2} \mu, \; \; \; \; \; \gamma^{(2)}=0, \; \; \; \; \; \mu^{(1)}_{AB} = \mu^{(2)}_{AB} = 0.
\end{equation}
Substituting these values to the results (\ref{Dltn1}) $-$ (\ref{Dltn2}) for $n>1$ and (\ref{Dlt12}) for $n=1$, it is straightforwardly proved that $[\Delta^{ab}]^{(i)} = 0$ for $i=1,2$ for a general $n$. Since $\Delta^{ab}$ vanishes  for $n> 1$, the boundary stress tensor becomes
\begin{align}
T_{ab} = -\frac{1}{8 \pi G} \frac{1}{r^{n-1}} \bigg(\frac{n}{2r^2} h^{(1)}_{ab} + \frac{1}{r^2} \gamma^{(1)}h^{(0)}_{ab}  + \frac{n}{(n-1)} \alpha \mu^{(0)}_{ab} + \frac{1}{(n-1)}\gamma^{(1)} \mu^{(0)}_{ab} + \frac{1}{(n-1)} D_{a} D_{b} \alpha  \bigg),
\end{align}
and the conserved charge is directly obtained
\begin{align}
Q[\xi^{t}] =& \frac{1}{8 \pi G} \int d^{n+1} x \sqrt{\gamma^{(0)}} u^{(0)t} T^{(1)}_{tt} \xi^{(0)t}, \nonumber\\
=& \frac{1}{8 \pi G} r^{n+1} A_{n+1} (-1) \bigg( - \frac{(n+1)}{2 r^{n+1}} \mu \bigg), \nonumber\\
=& \frac{(n+1)}{(n+2)} M.
\end{align}
For $n=1$ the boundary stress tensor has the form
\begin{align}
T_{ab} = - \frac{1}{8 \pi G} \bigg( \frac{1}{2r^2} h^{(1)}_{ab} - \frac{1}{2} \mu^{(1)}_{ab}\bigg),
\end{align}
and the conserved charge is
\begin{align}
Q[\xi^{t}] = \frac{1}{8 \pi G} \int^{2 \pi}_{0} d \varphi \int^{\pi}_{0} d \theta \; r^2 \sin{\theta} (-1) \bigg( - \frac{1}{r^{2}} \mu \bigg) = M,
\end{align}
which corresponds to {\cite{Myers:1986}}.

%----------------------------------------------
%----------------------------------------------
\section{Discussion}
\label{sec:Dscss}
%----------------------------------------------
%----------------------------------------------

We have computed the boundary stress tensor in $(n+3)$ dimensions associated with MM-counterterm in asymptotically flat static spacetime for cylindrical boundary surfaces as $r \rightarrow \infty$. We began with defining the most general form of the asymptotically static metric and then solved the decomposed Einstein equations and the MM-relation.  We found the MM-counterterm solution $\hat{K}_{ab}$ to be  uniquely determined for $n \geq 2$, but had ambiguities for $n=1$.

For $n=1$, at sub-leading
order these ambiguities can be nullified by choosing $\mu^{(1)}_{ab} = - 2 \alpha \mu^{(0)}_{ab}$ and $D_{a} D_{b} \alpha = - \alpha \mu^{(0)}_{ab} - \gamma^{(1)} \mu^{(0)}_{ab}$.   The quantity $\Delta_{ab}$ consequently vanishes.  At   sub-sub-leading order we found that while these ambiguities in $\hat{K}_{ab}$ cannot ensure that the resultant contribution to $\Delta_{ab}$ vanishes, as displayed in (\ref{Dlt11}) $-$ (\ref{Dlt12}),  they can
be chosen to ensure that $\Delta_{ab}$ does not contribute to the conserved charge.  These results are similar to those obtained for  the hyperbolic case \cite{Mann:2008}, which has manifestly vanishing $[\Delta^{ab}]^{(0)}$ and $[\Delta^{ab}]^{(1)}$, but non-vanishing $[\Delta^{ab}]^{(2)}$.  The stress-energy tensor is conserved provided (\ref{dvT12}) holds, which can be obtained by applying (\ref{cnstcmb11}) $-$ (\ref{cnstcmb13}).

For $n\geq 2$ we find at both sub-leading and sub-sub-leading orders that $\hat{K}_{ab}$ is determined. The quantity
$\Delta^{ab}$ turns out to be non-zero,  as shown in (\ref{Dltn0}) $-$ (\ref{Dltn2}).   This result indicates that the boundary stress tensor should be $T_{ab}$ in (\ref{BStnsr}) not $T^{\pi}_{ab}$ in (\ref{BStnsr_pi}). However we find that the contribution from $\Delta^{ab}$ does not contribute to the conserved charge (see (\ref{Q1},\ref{Q2}));  only $T^{\pi}_{ab}$ produces conserved charges and so the form of the conserved quantity formula (\ref{cnsvdcg}) is still valid. We also investigated the divergence of the boundary stress tensor, and found that  $\mathcal{D}^{a} T_{ab} = 0$ in appendix F.

We demonstrated   for  a static black hole in $(n+3)$-dimensional static spacetime  that $\Delta^{ab}$ manifestly is zero, and obtained the conserved charge from the boundary stress tensor. This agrees with the ADM mass, demonstrating that the boundary stress tensor with MM-counterterm is also applicable using cylindrical boundary conditions.

As  mentioned in section 2.2, there are some distinguishing properties between the hyperbolic boundary case and cylindrical boundary case. In the hyperbolic case, all components of the induced metric $h_{ab}$ can be expanded in the same order in  $r$, and so  are covariant under the variation. However in the cylindrical case  the induced metric $h_{ab}$ is again decomposed into $tt$- and angular components; these components have expansions to  different orders in $r$. They do not covariantly transform,  thereby not permitting inference of results from the hyperbolic case to the  cylindrical case. Furthermore, these two boundary conditions yield different solutions from the decomposed Einstein equations. In the hyperbolic case, the sub-sub-leading order of the Ricci tensor and the trace of the sub-sub-leading order of $h_{ab}$ become zero, so they in turn affect the sub-sub-leading order of $\hat{K}_{ab}$ and subsequently imply that the sub-sub-leading order of $\Delta^{ab}$ is zero. By contrast, in the cylindrical case the sub-sub-leading order of the Ricci tensor and the trace of the sub-sub-leading order of $h_{ab}$ are not zero;  they partly contribute to the sub-sub-leading order of $\Delta^{ab}$, rendering it nonzero. Despite these  differing properties between two boundary conditions, we found that the MM-counterterm is still valid yielding different descriptions of the boundary stress tensor in more than 4 dimensions for the respective cases.

\section*{\bf Acknowledgements}
M. Park would like to thank Keith Copsey for useful discussions. This work was supported in part by the Natural Sciences and Engineering Research Council of Canada.

%%%%%%%%%%%%%%%%%%%%%%%%%%%%%%%%%%%%%%%%%%%%
% Appendices
%%%%%%%%%%%%%%%%%%%%%%%%%%%%%%%%%%%%%%%%%%%%
\appendix

%----------------------------------------------
%----------------------------------------------
\section{Variation of the Action and the form of ${\Delta^{ab}}$}
\label{sec:VariatonAct}
%----------------------------------------------
%----------------------------------------------

The variation of the action (\ref{vrtnAC}) with respect to $h_{ab}$ is
\begin{equation}
\delta S = \frac{1}{ 16 \pi G} \int_{\partial {\cal{M}}} \sqrt{-h} \left[ \left( - \pi^{ab} - h^{ab} \hat{K} +2 \hat{K}^{ab} \right) \delta h_{ab} - 2 h^{ab} \delta \hat{K}_{ab} \right] \label{vrtnACappdx1}
\end{equation}
and to express $\delta \hat{K}_{ab}$ as a form of $\delta h_{ab}$, we take a derivation of (\ref{MMrltn}) with respect to $h_{ab}$
\begin{equation}
\delta {\cal{R}}_{cd} = \delta \hat{K}_{ab} L_{cd}^{\; \; \; \;ab} + \bigg( \hat{K}_{cd} \hat{K}_{mn} - \hat{K}_{cm} \hat{K}_{nd} \bigg) \delta h^{mn} \label{vrtRcc}
\end{equation}
where $L_{cd}^{\; \; \; \;ab}$ implies
\begin{equation}
L_{cd}^{\; \; \; \;ab} = h^{ab} \hat{K}_{cd} + \delta^{a}_{c} \delta^{b}_{d} \hat{K} - \delta^{a}_{c} \hat{K}^{b}_{d} - \delta^{b}_{c} \hat{K}^{a}_{d}. \label{expsL}
\end{equation}
Using the identity
\begin{equation}
(L^{-1})_{ab}^{\; \; \; mn} (L)_{mn}^{\; \; \; cd} = \delta^{c}_{a} \delta^{d}_{b}, \label{IdntL}
\end{equation}
(\ref{vrtRcc}) is changed to
\begin{equation}
\delta \hat{K}_{ab} = (L^{-1})_{ab}^{\; \; \; \; cd} \left[ \delta {\cal{R}}_{cd} + \bigg( \hat{K}_{cd} \hat{K}^{kl} - \hat{K}^{k}_{c} \hat{K}^{l}_{d} \bigg) \delta h_{kl} \right],
\end{equation}
and then (\ref{vrtnACappdx1}) is rearranged to
\begin{equation}
\delta S = \frac{1}{16 \pi G} \int_{ \partial {\cal{M}}} \sqrt{-h} \left[ \bigg( - \pi^{ab} + \hat{\pi}^{ab} + \hat{K}^{ab} - 2 \tilde{L}^{cd} \bigg( \hat{K}_{cd} \hat{K}^{ab} - \hat{K}_{c}^{a} \hat{K}_{d}^{b} \bigg)  \bigg) \delta h_{ab} - 2 \tilde{L}^{ab} \delta \mathcal{R}_{ab} \right] \label{vrtnACappdx2}
\end{equation}
where $\hat{\pi}^{ab} = \hat{K}^{ab} - h^{ab} \hat{K}$, and $\tilde{L}^{ab}$ is defined in (\ref{tldL}). If using the fact that
\begin{equation}
\delta R_{ab} = -\frac{1}{2} h^{kl} D_{a} D_{b} \delta h_{kl} - \frac{1}{2} h^{kl} D_{k} D_{l} \delta h_{ab} + h^{kl} D_{k} D_{(a} \delta h_{b)l}
\end{equation}
and doing integration by parts, (\ref{vrtnACappdx2}) takes
\begin{eqnarray}
\delta S = \frac{1}{16 \pi G} \int_{\partial {\cal{M}}} \sqrt{-h} \bigg[ - \pi^{ab} + \hat{\pi}^{ab} + \hat{K}^{ab} - 2 \tilde{L}^{cd} \bigg( \hat{K}_{cd} \hat{K}^{ab} - \hat{K}^{a}_{c} \hat{K}^{b}_{d} \bigg)  \nonumber\\
+ D^2 \tilde{L}^{ab} + h^{ab} D_{k} D_{l} \tilde{L}^{kl} - D_{k} \bigg( D^{a} \tilde{L}^{kb} + D^{b} \tilde{L}^{ka} \bigg) \bigg] \delta h_{ab}.
\end{eqnarray}
From the definition of the boundary stress tensor for the asymptotically flat spacetimes, $T_{ab}$ in (\ref{BStnsr}) is derived
\begin{displaymath}
T^{ab} = - \frac{2}{\sqrt{-h}} \frac{\delta S}{\delta h_{ab}} = \frac{1}{8 \pi G} \bigg( \pi^{ab} - \hat{\pi}^{ab} - \Delta^{ab} \bigg),
\end{displaymath}
and $\Delta^{ab}$ indicates
\begin{equation}
\Delta^{ab} = \hat{K}^{ab} - 2 \tilde{L}^{cd} \bigg( \hat{K}_{cd} \hat{K}^{ab} - \hat{K}^{a}_{c} \hat{K}^{b}_{d} \bigg)
+ {\cal{D}}^2 \tilde{L}^{ab} + h^{ab} {\cal{D}}_{k} {\cal{D}}_{l} \tilde{L}^{kl} - {\cal{D}}_{k} \bigg( {\cal{D}}^{a} \tilde{L}^{kb} + {\cal{D}}^{b} \tilde{L}^{ka} \bigg). \nonumber
\end{equation}

%----------------------------------------------
%----------------------------------------------
\section{Decomposed Einstein Equations}
\label{sec:DcmpsEE}
%----------------------------------------------
%----------------------------------------------

%Bringing the decomposed Einstein equation (3.1),
%\begin{displaymath}
%\perp(R_{ab}) = {\cal{R}}_{ab} + {\cal{D}}_{a} a_{b} - a_{a}a_{b} - \pounds_{n}K_{ab} - KK_{ab} + 2 K_{a}^{\; c}K_{cb}
%\end{displaymath}
%the third term in the right side expands
%\begin{eqnarray}
%\text{\pounds}_{n} K_{ab} &=& n^{c} \nabla_{c} K_{ab}  + \nabla_{a} n^{c} \; K_{cb} + \nabla_{b} n^{c} \; K_{ac} \nonumber\\
%&=& n^{c} \nabla_{c} K_{ab} + (K^{\; c}_{a} + n_{a} a^{c}) K_{cb} + (K^{\; c}_{b} + n_{b} a^{c}) K_{ac} \nonumber\\
%&=& n^{c} \nabla_{c} K_{ab} + 2 K^{\; c}_{a} K_{cb}
%\end{eqnarray}
%The Ricci scalars on the hypersurface from (3.3) expands
%\begin{align}
%{\cal{R}} =& \bigg({\cal{R}}^{(0)}_{ab} + \frac{1}{r^n} {\cal{R}}^{(1)}_{ab} + \frac{1}{r^{n+1}} {\cal{R}}^{(2)}_{ab} + \cdots \bigg) \bigg( h^{(0)ab} - \frac{1}{r^n} h^{(1)ab} - \frac{1}{r^{n+1}} h^{(2)ab} - \cdots \bigg) \nonumber\\
%=& {\cal{R}}^{(0)} + \frac{1}{r^n} \bigg({\cal{R}}^{(0)} -{\cal{R}}^{(0)}_{ab} h^{(1)ab} \bigg) + \frac{1}{r^{n+1}} \bigg({\cal{R}}^{(2)} - {\cal{R}}^{(0)}_{ab} h^{(2)ab} \bigg) - \frac{1}{r^{2n}} {\mathcal{R}}^{(1)}_{ab} h^{(1)ab} + \cdots
%\end{align}

%----------------------------------------------
\subsection{$n \geq 2$ Case}
%----------------------------------------------

In the asymptotically flat spacetime, which is described by the metric (\ref{Mtr}), the extrinsic curvature is calculated
\begin{equation}
K_{ab} = r \mu^{(0)}_{ab} + \frac{1}{r^{n-1}} \bigg(\mu^{(1)}_{ab} - \alpha \mu^{(0)}_{ab} - \frac{n}{2 r^2} h^{(1)}_{ab} \bigg) + \frac{1}{r^{n}} \bigg( \mu^{(2)}_{ab} - \frac{(n+1)}{2 r^2} h^{(2)}_{ab} \bigg) + {\cal{O}} \bigg( \frac{1}{r^{n+3}} \bigg),
\end{equation}
and taking the trace of it yields
\begin{eqnarray}
K = \frac{(n+1)}{r} - \frac{1}{r^{n+1}} \bigg( \alpha (n+1) + \frac{n}{2} h^{(1)} \bigg) - \frac{1}{r^{n+2}} \frac{(n+1)}{2} h^{(2)}  + {\cal{O}} \bigg( \frac{1}{r^{n+3}} \bigg)
\end{eqnarray}
where $K = K_{ab} h^{ab}$ and $h^{(m)}=h^{(m)}_{ab} h^{(0)ab}$ for $m=1,2$. The acceleration becomes
\begin{equation}
a_{a} =  \bigg( 0, - \frac{1}{r^n} D_{a} \alpha \bigg).
\end{equation}
The first decomposed Einstein equation ({\ref{DCEE1}}) is expanded
\begin{align}
0 &= {\cal{R}}^{(0)}_{ab} - n \mu^{(0)}_{ab} + \frac{1}{r^n} \left[ {\cal{R}}^{(1)}_{ab} - \bigg( n \mu^{(1)}_{ab} - n \alpha \mu^{(0)}_{ab} - \frac{n}{2} h^{(1)}\mu^{(0)}_{ab} + D_{a} D_{b} \alpha \bigg) \right] \nonumber\\
&+ \frac{1}{r^{n+1}} \left[ {\cal{R}}^{(2)}_{ab} - \bigg( n \mu^{(2)}_{ab} - \frac{(n+1)}{2} h^{(2)} \mu^{(0)}_{ab} + \frac{(n+1)}{2 r^2} h^{(2)}_{ab} \bigg) \right] +  {\cal{O}} \bigg( \frac{1}{r^{n+3}} \bigg), \label{DCEE1soln}
\end{align}
the second one (\ref{DCEE2}) takes
\begin{align}
0 &= \frac{1}{r^{n+1}} \left[ D_{a} \gamma^{(1)} + n D_{a} \alpha - \frac{n}{2} \bigg( D^{b} h^{(1)}_{ab} - D_{a} h^{(1)} \bigg) \right] \nonumber\\
& + \frac{1}{r^{n+2}} \left[ D_{a} \gamma^{(2)} - \frac{(n+1)}{2} \bigg( D^{b} h^{(2)}_{ab} - D_{a} h^{(2)} \bigg) \right] +  {\cal{O}} \bigg( \frac{1}{r^{n+3}} \bigg), \label{DCEE2soln}
\end{align}
and the last ({\ref{DCEE3}}) gives
\begin{align}
& 0 = {\cal{R}}^{(0)} - \frac{n(n+1)}{r^2} + \frac{1}{r^n} \bigg[ {\cal{R}}^{(1)} - \bigg( - {\frac{2 n (n+1)}{r^2} \alpha} - \frac{n(n+1)}{r^2} h^{(1)} + \frac{2n}{r^2}\mu^{(1)} \bigg)   \bigg] \nonumber\\
& + \frac{1}{r^{n+1}} \bigg[ {\cal{R}}^{(2)} - \bigg( \frac{(2n+1)}{r^2} \mu^{(2)} - \frac{(n+1)^2}{r^2} h^{(2)} \bigg)  \bigg]  +  {\cal{O}} \bigg( \frac{1}{r^{n+3}} \bigg). \label{DCEE3soln}
\end{align}
Note that the asymptotic expansion of ${\mathcal{R}^{(m)}_{ab}}$ is defined as
\begin{equation}
{\cal{R}}^{(m)}_{ab} = \frac{1}{2} \bigg( D^{c}D_{a} h^{(m)}_{cb} + D^{c} D_{b} h^{(m)}_{ac} - D^{c} D_{c} h^{(m)}_{ab} - D_{a} D_{b} h^{(m)} \bigg) \label{AspExRicci}
\end{equation}
where $m=1,2$, and $\mathcal{R}^{(m)}$ is a trace of (\ref{AspExRicci})
\begin{equation}
\mathcal{R}^{(m)} = h^{(0)ab} {\mathcal{R}}^{(m)}_{ab}.\label{AspExRicciS}
\end{equation}
As the solutions on the decomposed Einstein equations have to be consistent each other, we first compare ${\mathcal{R}}$'s, one from contracting ${\mathcal{R}}_{ab}$ in ({\ref{DCEE1soln}}) with $h^{(0)ab}$ and the other from (\ref{DCEE3soln}), then we get
\begin{align}
\label{CNSTcdt1forn} {\cal{R}}^{(1)} &= 2 D^2 \alpha = \frac{2 n}{r^2} \mu^{(1)} - \frac{2 n(n+1)}{r^2} \alpha - \frac{n(n+1)}{r^2} h^{(1)}, \\
\label{CNSTcdt2forn} {\cal{R}}^{(2)} &= \frac{n}{2 r^2} h^{(2)} = \frac{n}{(n+2)r^2} \mu^{(2)} = - \frac{2}{r^2} \gamma^{(2)}.
\end{align}
Now, taking the covariant derivative $D^a$ to ({\ref{DCEE2soln}}) leads the expression for Ricci scalar via (\ref{AspExRicci}) $-$ (\ref{AspExRicciS}) and this Ricci scalar quantity is satisfied with ({\ref{CNSTcdt1forn}}) $-$ ({\ref{CNSTcdt2forn}}) if
\begin{align}
\label{CNSTcdt3forn} D^2 \gamma^{(1)} = 0,  \\
\label{CNSTcdt4forn} D^2 \gamma^{(2)} = - \frac{(n+1)}{r^2} \gamma^{(2)}, \; \; D^2 \mu^{(2)} = - \frac{(n+1)}{r^2} \mu^{(2)}, \; \; D^2 h^{(2)} = - \frac{(n+1)}{r^2} h^{(2)}.
\end{align}

%----------------------------------------------
\subsection{$n = 1$ Case}
%----------------------------------------------

As mentioned in Sec 3., the $n=1$ case needs to be separately dealt with from the case with general $n$, because the sub-sub-leading order is expressed not only by the sub-sub-leading order quantities, but also by the combination of the sub-leading order values.

With the metric (\ref{Mtr}) having $n=1$, the extrinsic curvature at the boundary of the spacetime yields
\begin{equation}
            K_{ab} = r \mu^{(0)}_{ab} + \bigg(\mu^{(1)}_{ab} - \alpha \mu^{(0)}_{ab} - \frac{1}{2 r^2} h^{(1)}_{ab} \bigg) + \frac{1}{r} \bigg( \mu^{(2)}_{ab} - \alpha \mu^{(1)}_{ab} + \frac{\alpha}{2 r^2} h^{(1)}_{ab} - \frac{1}{r^2} h^{(2)}_{ab} \bigg) + {\cal{O}} \bigg( \frac{1}{r^{2}} \bigg),
\end{equation}
and its trace is
\begin{eqnarray}
K = \frac{2}{r} - \frac{1}{r^2} \bigg( 2 \alpha + \frac{1}{2} h^{(1)} \bigg) + \frac{1}{r^3} \bigg(- h^{(2)} + \frac{\alpha}{2} h^{(1)} + \frac{1}{2} h^{(1)}_{ab} h^{(1)ab} \bigg) + {\cal{O}} \bigg( \frac{1}{r^{4}} \bigg).
\end{eqnarray}
The acceleration $a_{a}$ becomes
\begin{equation}
a_{a} =  \bigg( 0, - \frac{1}{r} D_{a} \alpha + \frac{1}{r^{2}} \alpha D_{a} \alpha \bigg).
\end{equation}
Solving the decomposed Einstein equations as the previous section, (\ref{DCEE1}) yields
\begin{align}
        0 =& {\cal{R}}^{(0)}_{ab} - \mu^{(0)}_{ab} + \frac{1}{r} \left[ {\cal{R}}^{(1)}_{ab} - \bigg(\mu^{(1)}_{ab} - \alpha \mu^{(0)}_{ab} - \frac{1}{2} h^{(1)} \mu^{(0)}_{ab} + D_{a} D_{b} \alpha \bigg) \right] \nonumber\\
        &+ \frac{1}{r^{2}} \bigg[ {\cal{R}}^{(2)}_{ab} - \bigg( \mu^{(2)}_{ab}+ \frac{1}{r^2} h^{(2)}_{ab} - h^{(2)} \mu^{(0)}_{ab} - \alpha \mu^{(1)}_{ab} - \frac{\alpha}{r^2} h^{(1)}_{ab} + \alpha h^{(1)} \mu^{(0)}_{ab} \nonumber\\
        &- \frac{1}{2} h^{(1)} \mu^{(1)}_{ab} + \frac{1}{4r^2} h^{(1)} h^{(1)}_{ab} + \frac{1}{2} h^{(1)}_{cd} h^{(1)cd} \mu^{(0)}_{ab}- \frac{1}{2r^2} h^{(1)e}_{\; \; a} h^{(1)}_{e b} - \alpha D_{a} D_{b} \alpha \nonumber\\
        &- \frac{1}{2} (D_{a}h^{(1)}_{bd} + D_{b} h^{(1)}_{ad} - D_{d}h^{(1)}_{ab})D^{d} \alpha  \bigg) \bigg] +  {\cal{O}} \bigg( \frac{1}{r^{3}} \bigg), \label{DCEE1sol1}
\end{align}
(\ref{DCEE2}) takes
\begin{align}
    0 &= \frac{1}{r^2}\left[ D_{a} \gamma^{(1)} + D_{a} \alpha - \frac{1}{2} \bigg( D^{b} h^{(1)}_{ab} - D_{a} h^{(1)} \bigg) \right] + \frac{1}{r^3} \bigg[ \bigg( D_{a} \gamma^{(2)} + \frac{\alpha}{2} (D^{b}h^{(1)}_{ab} - D_{a} h^{(1)}) \nonumber\\
    & + \frac{1}{2} h^{(1)c}_{\; \; a} D_{c} \alpha - \frac{1}{2} h^{(1)} D_{a} \alpha - \alpha D_{a} \gamma^{(1)} - 2 \gamma^{(1)} D_{a} \gamma^{(1)} - \frac{3}{4} h^{(1)cd} D_{a} h^{(1)}_{cd} + \frac{1}{2} h^{(1)}_{ea} D_{b} h^{(1)be} \nonumber\\
    & + \frac{1}{2} h^{(1)bc} D_{c} h^{(1)}_{ab} - \frac{1}{4} h^{(1)}_{ea} D^{e} h^{(1)} \bigg) - \bigg( D^{b} h^{(2)}_{ab} - D_{a} h^{(2)} \bigg) \bigg] +  {\cal{O}} \bigg( \frac{1}{r^{4}} \bigg),\label{DCEE2sol1}
\end{align}
and (\ref{DCEE3}) has
\begin{align}
        0 =& {\cal{R}}^{(0)} - \frac{2}{r^2} + \frac{1}{r} \left[ {\cal{R}}^{(1)} - \frac{1}{r^2} \bigg( -4 \alpha - 2 h^{(1)} + 2 \mu^{(1)}  \bigg)   \right] \nonumber\\
        & + \frac{1}{r^{2}} \bigg[ {\cal{R}}^{(2)} - \frac{1}{r^2}\bigg(  3 \mu^{(2)} - 4 h^{(2)} - 3 \alpha \mu^{(1)} + 4 \alpha h^{(1)} + 2 \alpha^2 - \frac{1}{2} h^{(1)} \mu^{(1)}  \nonumber\\
        & + \frac{1}{4} (h^{(1)})^2 + \frac{7}{4} h^{(1)}_{ab} h^{(1)ab} - r^2 \mu^{(1)}_{ab} h^{(1)ab} + r^2 h^{(1)ab} D_{a} D_{b} \alpha \bigg)  \bigg]  +  {\cal{O}} \bigg( \frac{1}{r^{4}} \bigg).  \label{DCEE3sol1}
\end{align}
For the sub-leading order, we hold the same consistency conditions (\ref{CNSTcdt1forn}) and (\ref{CNSTcdt3forn}) with $n=1$ from (\ref{DCEE1sol1}) $-$ (\ref{DCEE3sol1}), but (\ref{CNSTcdt1forn}) are especially illustrated as
\begin{equation}
\mathcal{R}^{(2)} = 2 D^2 \alpha = - \frac{4}{r^2} \alpha - \frac{4}{r^2} \gamma^{(1)},
\end{equation}
and from this, we can infer that
\begin{equation}
D_{a} D_{b} \alpha = - \alpha \mu^{(0)}_{ab} - \gamma^{(1)} \mu^{(0)}_{ab} \label{DDalph}
\end{equation}
which is useful later in calculating $\Delta^{ab}$. In addition, we have
\begin{equation}
        \mu^{(1)}_{ab} = - 2 \alpha \mu^{(0)}_{ab}, \; \; \; D_{a} \gamma^{(1)} = 0, \label{sptrnslt}
\end{equation}
by disposing of the supertranslation which requires that the magnetic part of the four dimensional Weyl tensor to be zero
\begin{align}
        &k_{ab} = h^{(1)}_{ab} + 2 \alpha r^2 \mu^{(0)}_{ab} - {2 \gamma^{(1)} u^{(0)}_{a} u^{(0)}_{b}}, \\
        &t_{ab} = \epsilon_{a}^{\; \; \; cd} D_{c} k_{bd} = 0 \label{weylmgn}.
\end{align}
Applying (\ref{CNSTcdt1forn}) and (\ref{sptrnslt}) to the sub-sub-leading order of (\ref{DCEE1sol1}) $-$ (\ref{DCEE3sol1}), we obtain
\begin{align}
\label{cnstcmb11}        &-\frac{1}{r^2} \mu^{(2)} - \frac{6}{r^2} \gamma^{(2)} + \frac{2}{r^2} \alpha^2 + \frac{8}{r^2} \alpha \gamma^{(1)} + \frac{2}{r^2} (\gamma^{(1)})^2 = 0, \\
\label{cnstcmb12}        &-\frac{1}{r^2} \mu^{(2)} - \frac{8}{r^2} \gamma^{(2)} + \frac{6}{r^2} \alpha \gamma^{(1)} + \frac{2}{r^2} (\gamma^{(1)})^2 - D^2 \gamma^{(2)} + 2 D_{a} \alpha D^{a} \alpha = 0, \\
\label{cnstcmb13}        &D^2 \gamma^{(2)} + \frac{2}{r^2} \gamma^{(2)} + \frac{2}{r^2} \alpha \gamma^{(1)} + \frac{2}{r^2} \alpha^2 - 2 D_{a} \alpha D^{a} \alpha = 0.
\end{align}

%----------------------------------------------
%----------------------------------------------
\section{Exact Solution of ${\hat{Q}}_{ab}$}
\label{sec:ExtSlt}
%----------------------------------------------
%----------------------------------------------

The MM-counterterm $\hat{K}_{ab}$ is changed to $\hat{Q}_{ab}$ in (\ref{MMcntQ}), and then the relation (\ref{MMrltn}) is rewritten as
\begin{equation}
r^2 (\hat{Q}_{ab} \hat{Q} - h^{cd} \hat{Q}_{ac} \hat{Q}_{bc}) = {\cal{R}}_{ab}.
\end{equation}

%----------------------------------------------
\subsection{$n \geq 2$ Case}
%----------------------------------------------

In order to solve the MM-relation, we need to rearrange ({\ref{MMrltnep1}}) $-$ ({\ref{MMrltnep3}}) on $\hat{Q}_{ab}$, and then for $n \geq 2$ case, $\hat{Q}_{ab}$ is uniquely determined as follows
\begin{align}
\hat{Q}^{(0)}_{ab} &= \mu^{(0)}_{ab}, \\
\label{htqRcn1} \hat{Q}^{(1)}_{ab} &= \frac{1}{(n-1)} \bigg[ {\mathcal{R}}^{(1)}_{ab} - \frac{r^2}{2n} {\mathcal{R}}^{(1)} \mu^{(0)}_{ab} - \frac{r^2}{n(n+1)} {\mathcal{R}}^{(1)}_{cd} u^{(0)c} u^{(0)d} \mu^{(0)}_{ab} + \frac{2}{(n+1)} u^{(0)}_{a} u^{(0)c} {\mathcal{R}}^{(1)}_{cb} + \frac{1}{2} \mu^{(1)} \mu^{(0)}_{ab} - \mu^{(1)}_{ab} \bigg], \\
\label{htqRcn2} \hat{Q}^{(2)}_{ab} &= \frac{1}{(n-1)} \bigg[ {\mathcal{R}}^{(2)}_{ab} - \frac{r^2}{2n} {\mathcal{R}}^{(2)} \mu^{(0)}_{ab} - \frac{r^2}{n(n+1)} {\mathcal{R}}^{(2)}_{cd} u^{(0)c} u^{(0)d} \mu^{(0)}_{ab} + \frac{2}{(n+1)} u^{(0)}_{a} u^{(0)c} {\mathcal{R}}^{(2)}_{cb} + {\frac{1}{2} \mu^{(2)}} \mu^{(0)}_{ab} - \mu^{(2)}_{ab} \bigg]
\end{align}
where $u_{a}$ is the timelike normal vector and $\mu_{ab}$ is the pull-back metric of $\mu_{AB}$ for $A,B = \theta_{1}, ..., \theta_{n+1}$ on $(n+1)$-dimensional spacelike hypersurface, and $\hat{Q}^{(m)}_{ab}$, $\mathcal{R}^{(m)}_{ab}$, $u^{(m)}_{a}$ or $\mu^{(m)}_{ab}$ for $m=0,1,$ and $2$ are lowered and raised by $h^{(0)}_{ab}$. As $\mathcal{R}^{(1)}_{ab}$ is constituted of the pull-back metric components $\mu^{(0)}_{ab}$ and $\mu^{(1)}_{ab}$, the first sub-leading order $\hat{Q}^{(1)}_{ab}$ has just angular components, since $\alpha$ is independent of time, $t$. Since $\mathcal{R}^{(2)}_{ab}$ has $tt$-component and angular components, $\hat{Q}^{(2)}_{ab}$ also is expressed by $tt$- and angular components. Plugging (\ref{DCEE1soln}) and (\ref{DCEE2soln}) into (\ref{htqRcn1}) and (\ref{htqRcn2}), we finally obtain (\ref{qab1frn}) and (\ref{qab2frn}) in section 3.

%----------------------------------------------
\subsection{$n = 1$ Case}
%----------------------------------------------

As seen in (\ref{MMrltnep2}) and (\ref{MMrltnep4}), when $n=1$, as the $\hat{Q}_{ab}$ does not show up in the MM-relation, it is not directly obtainable. However, we can still derive the trace, $\hat{Q}^{(i)}$, which is
\begin{align}
\label{trcQ1}   {\hat{Q}}^{(1)} =& \frac{1}{2} {\mathcal{R}}^{(1)} + \frac{1}{2r^2} \mu^{(1)}, \\
\label{trcQ2}   {\hat{Q}}^{(2)} =& \frac{1}{2} {\mathcal{R}}^{(2)} + \frac{1}{2} {\mathcal{R}}^{(2)}_{cd} u^{(0)c} u^{(0)d} + \frac{1}{2 r^2} \mu^{(2)} - \frac{r^2}{8} ({\mathcal{R}}^{(1)})^2 + \frac{1}{2} {\mathcal{R}}^{(1)}_{cd} h^{(1)cd} + \frac{r^2}{8} {\mathcal{R}}^{(1)cd} {\mathcal{R}}^{(1)}_{cd}   \nonumber\\
        &- \frac{1}{r^2} \mu^{(1)cd} {\mathcal{R}}^{(1)}_{cd} + {\frac{1}{8r^2} (\mu^{(1)})^2}\ - \frac{3}{8r^2} \mu^{(1)}_{cd} \mu^{(1)cd},
\end{align}
where $\hat{Q}^{(i)} = \hat{Q}^{(i)}_{ab} h^{(0)ab}$ for $i=1,2$, and the contracted with the timelike normal vectors, $\hat{Q}^{(i)}_{ab} u^{(0)a} u^{(0)b}$, which is
\begin{align}
\label{Qtt1}    \hat{Q}^{(1)}_{tt} &= 0, \\
\label{Qtt2}    \hat{Q}^{(2)}_{tt} &= \frac{1}{2} \mathcal{R}^{(2)}_{tt} = - \frac{1}{r^2} \gamma^{(2)} + \frac{2}{r^2} \alpha \gamma^{(1)}.
\end{align}
From these values, the forms of $\hat{Q}^{(1)}_{ab}$ can be inferred as follows
\begin{align}
\label{antzQ11}     \hat{Q}^{(1)}_{ab} =& \beta_1 {\mathcal{R}}^{(1)}_{ab} + r^2 \beta_2 {\mathcal{R}}^{(1)} \mu^{(0)}_{ab} + \lambda_1 \; \mu^{(1)}_{ab} + \lambda_2 \; \mu^{(1)} \mu^{(0)}_{ab},
\end{align}
where $\beta_1$, $\beta_2$, $\lambda_1$ and $\lambda_2$ are ambiguities, which are not fixed from (\ref{trcQ1}) and (\ref{Qtt1}), and restricted to $\beta_1 + 2 \beta_2 = \frac{1}{2}$, and $\lambda_1 + 2 \lambda_2 = \frac{1}{2}$. If applying (\ref{DDalph}) and (\ref{sptrnslt}) into (\ref{antzQ11}), we have
\begin{align}
\hat{Q}^{(1)}_{ab} &= \beta_1 (-\alpha \mu^{(0)}_{ab} - \gamma^{(1)} \mu^{(0)}_{ab} + D_a D_b \alpha) + 2 \beta_2 (- 2 \alpha - 2 \gamma^{(1)}) \mu^{(0)}_{ab} - 2 \alpha \lambda_1 \mu^{(0)}_{ab} - 4 \alpha \lambda_2 \mu^{(0)}_{ab}, \nonumber\\
&= \beta_1 (-2 \alpha  - 2 \gamma^{(1)} ) \mu^{(0)}_{ab} + 2 \beta_2 (- 2 \alpha - 2 \gamma^{(1)}) \mu^{(0)}_{ab} - 2 \alpha (\lambda_1 + 2 \lambda_2) \mu^{(0)}_{ab}, \nonumber\\
&= (\beta_1 + 2 \beta_2) (-2 \alpha - 2 \gamma^{(1)}) \mu^{(0)}_{ab} - 2 \alpha (\lambda_1 + 2 \lambda_2) \mu^{(0)}_{ab}, \nonumber\\
&= - 2 \alpha \mu^{(0)}_{ab} - \gamma^{(1)} \mu^{(0)}_{ab}.
\end{align}
As shown the above, regardless of the ambiguities we become to have the same expression for $\hat{Q}^{(1)}_{ab}$, and so we randomly fixed $\beta_1 = \frac{1}{2}$, $\beta_2=0$, $\lambda_1 = \frac{1}{2}$ and $\lambda_2=0$ in (\ref{qab1fr1}). For the sub-sub-leading term, we guess a general form of $\hat{Q}^{(2)}_{ab}$ from (\ref{trcQ2}) and (\ref{Qtt2})
\begin{align}
\label{antzQ12}        \hat{Q}^{(2)}_{ab} =& \kappa_1 {\mathcal{R}}^{(2)}_{ab}  + r^2 \kappa_2 \mathcal{R}^{(2)} \mu^{(0)}_{ab} + \kappa_3 {\mathcal{R}}^{(2)}_{cd} u^{(0)c} u^{(0)d} u^{(0)}_a u^{(0)}_b + r^2 \kappa_4 {\mathcal{R}}^{(2)}_{cd} u^{(0)c} u^{(0)d} \mu^{(0)}_{ab}  \nonumber\\
        &+ \chi_1 \; \mu^{(2)}_{ab} + \chi_2 \; \mu^{(0)} \mu^{(0)}_{ab} - \frac{r^2}{8} {\mathcal{R}}^{(1)} {\mathcal{R}}^{(1)}_{ab} + \frac{r^2}{8} {\mathcal{R}}^{(1)}_{ac} {\mathcal{R}}^{(1)c}_{ \; \; b} - \frac{1}{2} {\mathcal{R}}^{(1)}_{ac} \mu^{(1)c}_{\; \; \; b} \nonumber\\
        &+ \frac{1}{8} \mu^{(1)} \mu^{(1)}_{ab} - \frac{3}{8} \mu^{(1)}_{ac} \mu^{(1)c}_{\; \; \;b}.
\end{align}
where $\kappa_{i}$ for $i=1,..,4$ and $\chi_{j}$ for $j=1,2$ are ambiguities, which are related to
\begin{equation}
\kappa_1 + 2 \kappa_2 = \frac{1}{2}, \; \; \; \; -\kappa_3 + 2 \kappa_4 = \frac{1}{2}, \; \; \; \; \kappa_1 + \kappa_3 = \frac{1}{2}, \; \; \; \; \chi_1 + 2 \chi_2 = \frac{1}{2},
\end{equation}
and as we have seen, the ambiguities for the multiplication of the first orders are nullified due to (\ref{DDalph}) and (\ref{sptrnslt}). Expanding (\ref{antzQ12}), it yields
\begin{equation}
\hat{Q}^{(2)}_{ab} = \lambda_3 \mu^{(2)}_{ab} + \lambda_4 \mu^{(2)} \mu^{(0)}_{ab} - \frac{3}{2} \gamma^{(2)} \mu^{(0)}_{ab} - \frac{1}{r^2} \gamma^{(2)} u^{(0)}_{a} u^{(0)}_{b} - \frac{5}{2} \alpha^2 \mu^{(0)}_{ab} + \frac{2}{r^2} \alpha \gamma^{(1)} u^{(0)}_{a} u^{(0)}_{b}
\end{equation}
where redefined $ \lambda_3 = 2 \kappa_1 + \chi_1$ and $\lambda_4 = \chi_2 - \kappa_1$. The ambiguities $\lambda_3$ and $\lambda_4$ are determined at the end of the calculation of $[\Delta^{ab}]^{(2)}$ in a way that $[\Delta^{ab}]^{(2)}$ does not contribute to the conserved quantities, and it turns out $\lambda_3 = - \frac{1}{2}$ and $\lambda_4 = \frac{1}{2}$. Our solution on $\hat{Q}^{(2)}_{ab}$ is displayed in (\ref{qab2fr1}).

%----------------------------------------------
%----------------------------------------------
\section{Explicit Form of ${\tilde{L}}^{ab}$}
\label{sec:ExplctFrm}
%----------------------------------------------
%----------------------------------------------

$L_{ab}^{\; \; \; \; cd}$ is defined in (\ref{expsL}) and is a shorthand expression for convenience to deal with terms constituted of $\hat{K}_{ab}$'s. Our interest is to get ${\tilde{L}}^{ab}$ which is defined in (\ref{tldL}). Firstly, we expand $L_{mn}^{\; \; \; \; cd}$ and $(L^{-1})_{ab}^{\; \; \; \;mn}$ in order of $r$
\begin{equation}
L_{mn}^{\; \; \; \; cd} = {L^{(0)}}_{mn}^{\; \; \; \; cd} + \frac{1}{r^n}{L^{(1)}}_{mn}^{\; \; \; cd} + \frac{1}{r^{n+1}}{L^{(2)}}_{mn}^{\; \; \; cd} + \cdots,
\end{equation}
\begin{equation}
{(L^{-1})}_{ab}^{\; \; \; mn} = {(L^{-1})^{(0)}}_{ab}^{\; \; \; mn} + \frac{1}{r^n} {(L^{-1})^{(1)}}_{ab}^{\; \; \; mn} + \frac{1}{r^{n+1}} {(L^{-1})^{(2)}}_{ab}^{\; \; \; mn} + \cdots.
\end{equation}
Plugging them into the identity relation (\ref{IdntL}), the relation is satisfied if
\begin{equation}
{(L^{-1})}_{ij}^{(0)\; kl} {L^{(0)}}_{kl}^{\; \; \; mn} = \delta^{m}_{i} \delta^{n}_{j} \label{IdntL0}
\end{equation}
and
\begin{align}
\label{IdntL1} {(L^{-1})}_{ij}^{(1)\;pq} =& - {(L^{-1})}_{ij}^{(0)\; kl} {L^{(1)}}_{kl}^{\; \; \; mn} {(L^{-1})}_{mn}^{(0)\; pq}, \\
\label{IdntL2} (n \geq 2), \; \; {(L^{-1})}_{ij}^{(2)\;pq} =& - {(L^{-1})}_{ij}^{(0)\; kl} {L^{(2)}}_{kl}^{\; \; \; mn} {(L^{-1})}_{mn}^{(0)\; pq}, \\
\label{IdntL3} (n = 1), \; \; {(L^{-1})}_{ij}^{(2)\;pq} =& - {(L^{-1})}_{ij}^{(0)\; kl} {L^{(2)}}_{kl}^{\; \; \; mn} {(L^{-1})}_{mn}^{(0)\; pq} - {(L^{-1})}_{ij}^{(1)\; kl} {L^{(1)}}_{kl}^{\; \; \; mn} {(L^{-1})}_{mn}^{(0)\; pq},
\end{align}
where ${L^{(i)}}_{kl}^{\; \; \; mn}$ can be directly red from (\ref{expsL})
\begin{align}
        \label{Lijkl0} {L^{(0)}}_{ij}^{\; \;kl} =& r h^{(0)kl} \mu^{(0)}_{ij} + \frac{(n+1)}{r} \delta^{k}_{i} \delta^{l}_{j} - \frac{2}{r} \delta^{k}_{(i} \mu^{(0)l}_{\; \; j)}, \\
        \label{Lijkl1} {L^{(1)}}_{ij}^{\; \;kl} =& r \bigg( h^{(0)kl} \hat{Q}^{(1)}_{ij} + \hat{Q}^{(1)} \delta^{k}_{i} \delta^{l}_{j} -2  \delta^{k}_{(i} \hat{Q}^{(1)l}_{\; \; j)} - h^{(1)kl} \mu^{(0)}_{ij} \bigg) + \frac{1}{r} \bigg( 2 \delta^{k}_{(i} \mu^{(1)l}_{\; \; j)} - \mu^{(1)} \delta^{k}_{i} \delta^{l}_{j} \bigg), \\
        \label{Lijkl2n} (n \geq 2), \; \; {L^{(2)}}_{ij}^{\; \;kl} =& r \bigg( h^{(0)kl} \hat{Q}^{(2)}_{ij} + \hat{Q}^{(2)} \delta^{k}_{i} \delta^{l}_{j} - 2 \delta^{k}_{(i} \hat{Q}^{(2)l}_{\; \; j)} - h^{(2)kl} \mu^{(0)}_{ij} \bigg) + \frac{1}{r} \bigg( 2 \delta^{k}_{(i} \mu^{(2)l}_{\; \; j)} - \mu^{(2)} \delta^{k}_{i} \delta^{l}_{j} \bigg), \\
        \label{Lijkl21}(n = 1), \; \; {L^{(2)}}_{ij}^{\; \;kl} =& r \bigg( h^{(0)kl} \hat{Q}^{(2)}_{ij} + \hat{Q}^{(2)} \delta^{k}_{i} \delta^{l}_{j} - 2 \delta^{k}_{(i} \hat{Q}^{(2)l}_{\; \; j)} - h^{(2)kl} \mu^{(0)}_{ij} - h^{(1)kl} \hat{Q}^{(1)}_{ij} + 2 \delta^{k}_{(i} \hat{Q}^{(1)}_{j)m} h^{(1)ml} \nonumber\\
        &- \hat{Q}^{(1)}_{ab} h^{(1)ab} \delta^{k}_{i} \delta^{l}_{j} + h^{(1)km} h^{(1)l}_{\; \; m} \mu^{(0)}_{ij} \bigg) + \frac{1}{r} \bigg(2 \delta^{k}_{(i} \mu^{(2)l}_{\; \; j)} - \mu^{(2)} \delta^{k}_{i} \delta^{l}_{j} + \mu^{(1)ab} \mu^{(1)}_{ab} \delta^{k}_{i} \delta^{l}_{j} \nonumber\\
        &- 2 \delta^{k}_{(i} \mu^{(1)}_{j) m} \mu^{(1)ml} \bigg).
\end{align}
In which, when $n=1$ the sub-sub-leading order term is separately considered from one of general $n$, because the combination of the sub-leading order terms contributes to the sub-sub-leading order. From the definition (\ref{tldL}), $\tilde{L}^{ab}$ is expanded
\begin{align}
\tilde{L}^{ab} & = \bigg( h^{(0)mn} - \frac{1}{r^n}h^{(1)mn} - \frac{1}{r^{n+1}} h^{(2)mn} + \cdots \bigg) \bigg({(L^{-1})^{(0)}}_{mn}^{\; \; \; ab} + \frac{1}{r^n} {(L^{-1})^{(1)}}_{mn}^{\; \; \; ab} + \frac{1}{r^{n+1}} {(L^{-1})^{(2)}}_{mn}^{\; \; \; ab} + \cdots \bigg) \nonumber\\
& = \tilde{L}^{(0)ab} + \frac{1}{r^n} \tilde{L}^{(1)ab} + \frac{1}{r^{n+1}} \tilde{L}^{(2)ab} + \cdots,
\end{align}
where the each order becomes
\begin{align}
        \label{tldL1}\tilde{L}^{(0)ab} &= h^{(0)mn} (L^{-1})^{(0)\;ab}_{mn}, \\
        \label{tldL2} \tilde{L}^{(1)ab} &= h^{(0)mn} (L^{-1})^{(1) \; ab}_{mn} - h^{(1)mn} (L^{-1})^{(0) \; ab}_{mn}, \\
        \label{tldL3} (n \geq 2), \; \; \tilde{L}^{(2)ab} &= h^{(0)mn} (L^{-1})^{(2) \; ab}_{mn} - h^{(2)mn} (L^{-1})^{(0) \; ab}_{mn}, \\
        \label{tldL4} (n = 1), \; \; \tilde{L}^{(2)ab} &= h^{(0)mn} (L^{-1})^{(2) \; ab}_{mn} - h^{(2)mn} (L^{-1})^{(0) \; ab}_{mn} \nonumber\\
        &\; \; \; \; - h^{(1)mn} (L^{-1})^{(1) \; ab}_{mn} + h^{(1)ml} h^{(1)n}_{\; l} (L^{-1})^{(0) \; ab}_{mn}.
\end{align}

%----------------------------------------------
\subsection{$n \geq 2$ Case}
%----------------------------------------------

The inverse of ${(L)^{(0)}}_{mn}^{\; \; \; \; cd} $ is
\begin{align}
    (L^{-1})^{(0)\; \;ab}_{ij} =& \frac{r}{(n+1)} \bigg( \delta^{a}_{i} \delta^{b}_{j} - \frac{r^2}{2n} \mu^{(0)}_{ij} h^{(0)ab} + \frac{1}{n} \delta^{a}_{i} \mu^{(0)b}_{\; \; j} + \frac{1}{n} \delta^{b}_{i} \mu^{(0)a}_{\; \; j} \nonumber\\
    &+ \frac{1}{n(n-1)} \mu^{(0)b}_{\; \; i} \mu^{(0)a}_{\; \; j} + \frac{1}{n(n-1)} \mu^{(0)a}_{\; \; i} \mu^{(0)b}_{\; \; j} - \frac{1}{n(n-1)} \mu^{(0)}_{ij} \mu^{(0)ab} \bigg),
\end{align}
and contracting ${(L)^{(0)}}_{mn}^{\; \; \; \; cd} $ with $h^{(0)ij}$, we simply get
\begin{equation}
(L^{-1})^{(0)ab} = \tilde{L}^{(0)ab} =  \frac{r}{n(n+1)} \bigg( \frac{(n-1)}{2} h^{(0)ab} + \frac{1}{r^2} \mu^{(0)ab} \bigg).
\end{equation}
Once ${(L^{-1})}_{ij}^{(0)\; ab}$ is calculated, we can subsequently obtain ${(L^{-1})}_{ij}^{(1)\;pq}$ and ${(L^{-1})}_{ij}^{(2)\;pq}$ from the relation (\ref{IdntL1}) $-$ (\ref{IdntL2}). Contracting them with $h^{(0)ij}$ and plugging into (\ref{tldL2}) $-$ (\ref{tldL3}), we have
\begin{align}
    \tilde{L}^{(1)ab} =& \frac{r}{n(n-1)} \bigg( r^2 \hat{Q}^{(1)ab} - \frac{r^2}{2} \hat{Q}^{(1)}h^{(0)ab} - \frac{2nr^2}{(n+1)^2} \hat{Q}^{(1)}u^{(0)a} u^{(0)b} + \frac{1}{2} \mu^{(1)} h^{(0)ab} \nonumber\\
    & + \frac{2n}{(n+1)^2} \mu^{(1)} u^{(0)a} u^{(0)b} - \frac{(n-1)^2}{2(n+1)} h^{(1)ab} - \frac{2n}{(n+1)r^2} \mu^{(1)ab} \bigg),
\end{align}
and
\begin{align}
    \widetilde{L}^{(2)ab} =& \frac{r}{n(n-1)} \bigg( r^2 \hat{Q}^{(2)ab} - \frac{r^2}{2} \hat{Q}^{(2)}h^{(0)ab} - \frac{2nr^2}{(n+1)^2} \hat{Q}^{(2)}u^{(0)a} u^{(0)b} -\frac{r^2}{(n+1)} \hat{Q}^{(2)tt} h^{(0)ab} \nonumber\\
    &+ \frac{2 r^2}{n(n+1)^2} \hat{Q}^{(2)tt} u^{(0)a} u^{(0)b} - \frac{(2n^2+n+1)r^2}{n(n+1)^2}u^{(0)a} \hat{Q}^{(2)b}_{\; \; t} - \frac{(2n^2+n+1)r^2}{n(n+1)^2} \hat{Q}^{(2)a}_{\; \; t} u^{(0)b}  \nonumber\\
    &+ \frac{1}{2} \mu^{(2)} h^{(0)ab} + \frac{2n}{(n+1)^2} \mu^{(2)} u^{(0)a} u^{(0)b} - \frac{(n-1)^2}{2(n+1)} h^{(2)ab} - \frac{2n}{(n+1)r^2} \mu^{(2)ab} \bigg).
\end{align}

%----------------------------------------------
\subsection{$n = 1$ Case}
%----------------------------------------------

In $4$-dimensional spacetime case, the identity relation (\ref{IdntL0}) with ${L^{(0)}}_{ij}^{\; \;kl}$ in (\ref{Lijkl0}) gives
\begin{equation}
(L^{-1})^{(0)ab} = \tilde{L}^{(0)ab}= \frac{1}{2r} \mu^{(0)ab}, \label{tldL04d}
\end{equation}
and from this, we can find
\begin{align}
    (L^{-1})^{(0)ab}_{ij} = \frac{r}{2} \bigg( \delta^{a}_{i} \delta^{b}_{j} + \frac{r^2}{2} \mu^{(0)}_{ij} u^{(0)a} u^{(0)b} \bigg).
\end{align}
Then, $\tilde{L}^{(i)ab}$ for $i=1,2$ are expanded as
\begin{align}
    \tilde{L}^{(1)ab} =& (L^{-1})^{(1)ab} - h^{(1)mn} (L^{-1})^{(0)ab}_{mn} \nonumber\\
    =& \frac{r}{2} \bigg( r^2 \hat{Q}^{(1)ab} - \hat{Q}^{(1)}\mu^{(0)ab} + \frac{1}{2r^2} \mu^{(1)} \mu^{(0)ab} - \frac{1}{r^2} \mu^{(1)ab} \bigg), \\
    \tilde{L}^{(2)ab} =& (L^{-1})^{(2)ab} - h^{(1)mn} (L^{-1})^{(1)ab}_{mn} - h^{(2)mn} (L^{-1})^{(0)ab}_{mn} + h^{(1)ml} h^{(1)n}_{\; \; l} (L^{-1})^{(0)ab}_{mn} \nonumber\\
    =& - \frac{r}{4} \bigg( 2 \hat{Q}^{(2)} \mu^{(0)ab} + r^2 \hat{Q}^{(2)}_{tt} h^{(0)ab} - 2 r^2 \hat{Q}^{(2)ab} + 2 r^2 u^{(0)a} \hat{Q}^{(2)b}_{\; \; t} - \frac{1}{r^2} \mu^{(2)} \mu^{(0)ab} + \frac{2}{r^2} \mu^{(2)ab} \nonumber\\
    & - 2\hat{Q}^{(1)}_{mn} \mu^{(1)mn} h^{(0)ab} - 2 r^4 (\hat{Q}^{(1)})^2 h^{(0)ab} + r^4 \hat{Q}^{(1)mn} \hat{Q}^{(1)}_{mn} h^{(0)ab} + 2 \hat{Q}^{(1)} \mu^{(1)} \mu^{(0)ab} + 3 r^4 \hat{Q}^{(1)} \hat{Q}^{(1)ab} \nonumber\\
    & - 3 \hat{Q}^{(1)} \mu^{(1)ab} - 2 r^4 \hat{Q}^{(1)ac} \hat{Q}^{(1)b}_{\; \; c} + 6 \hat{Q}^{(1)a}_{\; \; c} \mu^{(1)bc} - 2 r^2 \mu^{(1)} \hat{Q}^{(1)ab} - \frac{3 r^4}{2} (\hat{Q}^{(1)})^2 u^{(0)a} u^{(0)b}  \nonumber\\
    &- r^2 \hat{Q}^{(1)} \mu^{(1)} u^{(0)a} u^{(0)b} + \frac{2}{r^2} \mu^{(1)} \mu^{(1)ab} - \frac{1}{2} (\mu^{(1)})^2 h^{(0)ab} + \mu^{(1)mn} \mu^{(1)}_{mn} h^{(0)ab} - \frac{4}{r^2} \mu^{(1)ac} \mu^{(1)b}_{\; \; c} \bigg).
\end{align}

%----------------------------------------------
%----------------------------------------------
\section{Calculation of $\Delta^{ab}$}
\label{sec:ClcltDelta}
%----------------------------------------------
%----------------------------------------------

Lastly, in order to calculate $\Delta^{ab}$, we need to expand a covariant derivative as
\begin{align}
{\cal{D}}_{a} \tilde{L}^{kl} =& D^{(0)}_{a} \tilde{L}^{(0)kl} + \frac{1}{r^n} \bigg( D^{(0)}_{a} \tilde{L}^{(1)kl} + D^{(1)}_{a} \tilde{L}^{(0)kl} \bigg) + \frac{1}{r^{n+1}} \bigg(D^{(0)}_{a} \tilde{L}^{(2)kl} + D^{(2)}_{a} \tilde{L}^{(0)kl} \bigg) \nonumber\\
+ & \frac{1}{r^{2n}} D^{(1)}_{a} \tilde{L}^{(1)kl} + {\cal{O}} \bigg( \frac{1}{r^{n+2}} \bigg)
\end{align}
where $D^{(0)}_{a}$ is a covariant derivative compatible with $h^{(0)}_{ab}$, and $D^{(0)}_{a}$ and $D^{(i)}_{a}$ for $i=1,2$ are denoted as
\begin{eqnarray}
D^{(0)}_{a} \tilde{L}^{(i)kl} &=& \partial \tilde{L}^{(i)kl} + \Gamma^{(0)k}_{\; \; \; am} \tilde{L}^{(i)ml} + \Gamma^{(0)l}_{\; \; \; am} \tilde{L}^{(i)mk} \\
D^{(i)}_{a} \tilde{L}^{(0)kl} &=& \Gamma^{(i)k}_{\; \; \; am} \tilde{L}^{(0)ml} + \Gamma^{(i)l}_{\; \; \; am} \tilde{L}^{(0)mk} \\
D^{(1)}_{a} \tilde{L}^{(1)kl} &=& \Gamma^{(1)k}_{\; \; \; am} \tilde{L}^{(1)ml} + \Gamma^{(1)l}_{\; \; \; am} \tilde{L}^{(1)mk}.
\end{eqnarray}
In which, the connection is written
\begin{equation}
\Gamma^{(i)a}_{\; \; \; \; bc} = \frac{1}{2} h^{(0)ae} \bigg(D_{b}h^{(i)}_{ce} + D_{c}h^{(i)}_{be} -D_{e} h^{(i)}_{bc} \bigg)
\end{equation}
where $D^{(0)}_{a}$ is simply denoted as $D_{a}$.

To simplify the expression of $\Delta^{ab}$, we use the commutation relation on the derivative as follows
\begin{align}
D_{k} D_{l} \bigg( D^{k} D^{l} \alpha \bigg) =& D_{k} D^2 (D^{k} \alpha) = D_{k} [D^2 , D^{k}] \alpha + D^2 D^2 \alpha \nonumber\\
=& -D_{k} \bigg( {\cal{R}}^{(0)km\; \; l}_{\; \; \; \; \; \; \; \; \;m} D_{l} \alpha \bigg) + D^2 D^2 \alpha \nonumber\\
=& \frac{n}{r^2} D^2 \alpha + D^2 D^2 \alpha \\
D^2 \bigg( D^a D^b \alpha \bigg) =& [D^2 , D^a] D^{b} \alpha + D^a [D^2 , D^{b}] \alpha + D^a D^b (D^2 \alpha) \nonumber\\
=& - {\cal{R}}^{(0)e\; \; ca}_{\; \; \; \; \; \; \; c} D_{e} D^{b} \alpha - {\cal{R}}^{(0)eb\; \; a}_{\; \; \; \; \; \; \; c} D^c D_{e} \alpha - D^c \bigg({\cal{R}}^{(0)eb\; \; a}_{\; \; \; \; \; \; \; c} D_{e} \alpha \bigg) \nonumber\\
& - D^a \bigg( {\cal{R}}^{(0)e\; \; cb}_{\; \; \; \; \; \;c} D_e \alpha \bigg) + D^a D^b (D^2 \alpha) \nonumber\\
=& 2 {\cal{R}}^{(0)caeb} D_{c} D_{e} + \frac{2n}{r^2} D^a D^b \alpha + D^a D^b D^2 \alpha \\
D_{k} D^{a} (D^{k} D^{b} \alpha) =& [D_{k}, D^{a}] (D^k D^b \alpha) + D^a [D^2, D^b] \alpha + D^a D^b (D^2 \alpha) \nonumber\\
=& - {\cal{R}}^{(0)e\; \; ca}_{\; \; \; \; \; \;c} D_{e} D^{b} \alpha - {\cal{R}}^{(0)ebca} D_{c} D_{e} \alpha - D^a \bigg( {\cal{R}}^{(0)e\; \; ca}_{\; \; \; \; \; \;c} D_{e} \alpha \bigg) + D^a D^b (D^2 \alpha) \nonumber\\
=& {\cal{R}}^{(0)caeb} D_{c} D_{e} \alpha + \frac{2n}{r^2} D^a D^b \alpha + D^a D^b D^2 \alpha
\end{align}
where the commutation of two covariant derivatives acting on $w_c$ and $t_{ab}$ is
\begin{eqnarray}
&& [D_{a}, D_{b}] w_{c} = {\cal{R}}^{(0)\; \;d}_{\; abc} w_{d} \nonumber\\
&& [D_{a}, D_{b}] t_{c}^{\; d} = {\cal{R}}^{(0)\; \;e}_{\; abc} t_{e}^{\; d} + {\cal{R}}^{(0)\;d}_{\; ab \; \; e} t_{c}^{\; e}.
\end{eqnarray}

\subsection{$n \geq 2$}
    \begin{align}
    &\bigg[ \hat{K}^{ab} - 2 \tilde{L}^{cd}(\hat{K}_{cd} \hat{K}^{ab} - \hat{K}^{a}_{c} \hat{K}^{b}_{d}) \bigg]^{(1)} = \nonumber\\
    & \frac{r}{n(n-1)} \bigg[ \frac{n(n+1)}{(n-1)r^4} \alpha \mu^{(0)ab} + \frac{n(n+1)}{(n-1)r^4} \gamma^{(1)} \mu^{(0)ab} + \frac{n}{2r^4} \mu^{(1)} \mu^{(0)ab} + \frac{(n+1)}{(n-1)} D^a D^b \alpha  \bigg], \\
    & \bigg[\mathcal{D}^2 \tilde{L}^{ab} + h^{ab} \mathcal{D}_{k} \mathcal{D}_{l} \tilde{L}^{kl} - \mathcal{D}_{k} (\mathcal{D}^a \tilde{L}^{kb} + \mathcal{D}^b \tilde{L}^{ka}) \bigg]^{(1)} = \nonumber\\
    &\frac{r}{n(n-1)} \bigg[ - \frac{n(n+1)}{(n-1)r^4} \alpha \mu^{(0)ab} - \frac{n(n+1)}{(n-1)r^4} \gamma^{(1)} \mu^{(0)ab} - \frac{n}{2r^4} \mu^{(1)} \mu^{(0)ab} - \frac{(n+1)}{(n-1)} D^a D^b \alpha  \nonumber\\
    &+ D^a D^b \gamma^{(1)} + \frac{n^2}{(n+1)^2} D^2 \mu^{(1)} u^{(0)a} u^{(0)b} + \frac{2n}{(n+1)} D^2 \alpha u^{(0)a} u^{(0)b} \bigg],
    \end{align}

    \begin{align}
    & \bigg[\hat{K}^{ab} - 2 \tilde{L}^{cd}(\hat{K}_{cd} \hat{K}^{ab} - \hat{K}^{a}_{c} \hat{K}^{b}_{d}) \bigg]^{(2)} = \nonumber\\
    & \frac{r}{n(n-1)} \bigg[ - \frac{(n-1)}{2r^2} h^{(2)ab}  + \frac{(n^2+1)}{(n-1)} \mu^{(2)ab} + \frac{2(n^3 + n^2 + 3n + 1)}{n(n-1)(n+1)r^4} \gamma^{(2)} \mu^{(0)ab} \bigg], \\
    & \bigg[\mathcal{D}^2 \tilde{L}^{ab} + h^{ab} \mathcal{D}_{k} \mathcal{D}_{l} \tilde{L}^{kl} - \mathcal{D}_{k} (\mathcal{D}^a \tilde{L}^{kb} + \mathcal{D}^b \tilde{L}^{ka}) \bigg]^{(2)} = \nonumber\\
    &\frac{r}{n(n-1)} \bigg[\frac{(n-1)}{2r^2} h^{(2)ab} - \frac{(2n^2+n+1)}{(n-1)r^4} \mu^{(2)ab} - \frac{(3r^2 + 7n + 4)}{n(n-1)r^4} \gamma^{(2)} \mu^{(0)ab}   \nonumber\\
    & + \frac{(n^3 - 2n^2 -n-2)}{n(n-1)(n+1)} D^a D^b \gamma^{(2)} \bigg]
    \end{align}

\subsection{$n = 1$}
    \begin{align}
    &\bigg[\hat{K}^{ab} - 2 \tilde{L}^{cd}(\hat{K}_{cd} \hat{K}^{ab} - \hat{K}^{a}_{c} \hat{K}^{b}_{d}) \bigg]^{(1)} = 2r \bigg( \frac{1}{r^4} \alpha \mu^{(0)ab} + \frac{1}{r^4} \gamma^{(1)} \mu^{(0)ab} + D^a D^b \alpha \bigg), \\
    &\bigg[\mathcal{D}^2 \tilde{L}^{ab} + h^{ab} \mathcal{D}_{k} \mathcal{D}_{l} \tilde{L}^{kl} - \mathcal{D}_{k} (\mathcal{D}^a \tilde{L}^{kb} + \mathcal{D}^b \tilde{L}^{ka}) \bigg]^{(1)} = \frac{3r}{2}\bigg( - \frac{1}{r^4} \alpha \mu^{(0)ab} - \frac{1}{r^4} \gamma^{(1)} \mu^{(0)ab} - D^a D^b \alpha \bigg)
    \end{align}

    \begin{align}
    &\bigg[\hat{K}^{ab} - 2 \tilde{L}^{cd}(\hat{K}_{cd} \hat{K}^{ab} - \hat{K}^{a}_{c} \hat{K}^{b}_{d}) \bigg]^{(2)} = \frac{1}{r} \bigg( \gamma^{(2)} u^{(0)a} u^{(0)a} - 2 \alpha \gamma^{(1)} u^{(0)a} u^{(0)b} - \frac{3}{r^4} \mu^{(2)ab} \nonumber\\
    & + \frac{1}{r^4} \mu^{(2)} \mu^{(0)ab} + \frac{1}{2r^4} \gamma^{(2)} \mu^{(0)ab} - \frac{1}{r^4} \alpha \gamma^{(1)} \mu^{(0)ab} \bigg), \\
    &\bigg[\mathcal{D}^2 \tilde{L}^{ab} + h^{ab} \mathcal{D}_{k} \mathcal{D}_{l} \tilde{L}^{kl} - \mathcal{D}_{k} (\mathcal{D}^a \tilde{L}^{kb} + \mathcal{D}^b \tilde{L}^{ka}) \bigg]^{(2)} = \frac{1}{r} \bigg( - \gamma^{(2)} u^{(0)a} u^{(0)a} + 2 \alpha \gamma^{(1)} u^{(0)a} u^{(0)b} + \frac{1}{r^4} \mu^{(2)ab} \nonumber\\
    & - \frac{1}{2r^4} \mu^{(2)} \mu^{(0)ab} - \frac{1}{r^4} \gamma^{(2)} \mu^{(0)ab} + \frac{2}{r^4} \alpha \gamma^{(1)} \mu^{(0)ab} + \frac{1}{2r^4} \alpha^2 \mu^{(0)ab} + \frac{1}{2r^4} (\gamma^{(1)})^2 \mu^{(0)ab} - \frac{r^2}{4} D^a D^b \mu^{(2)} \nonumber\\
    & - \frac{r^2}{2} D^a D^b \gamma^{(2)} + \frac{1}{4} D^2 \mu^{(2)} \mu^{(0)ab} + \frac{3}{4} D^2 \gamma^{(2)} \mu^{(0)ab}- 4 r^2 D^a \alpha D^b \alpha  - 5 r^2 \alpha D^a D^b \alpha - 3 r^2 \gamma^{(1)} D^a D^b \alpha  \nonumber\\
    &- \frac{1}{2} D^e \alpha D_{e} \alpha \mu^{(0)ab} + \frac{11}{4} D^2 \alpha^2 \mu^{(0)ab} + \frac{13}{4} D^2 (\alpha \gamma^{(1)}) \mu^{(0)ab} - \frac{3r^2}{2} D^2 \alpha^2 u^{(0)a} u^{(0)b} - \frac{5r^2}{4} \gamma^{(1)} D^2 \alpha u^{(0)a} u^{(0)b} \bigg)
    \end{align}

%----------------------------------------------
%----------------------------------------------
\section{Divergence of the Boundary Stress Tensor, $\mathcal{D}^{a} T_{ab}$}
%----------------------------------------------
%----------------------------------------------

In {\cite{Mann:2006}}, it was argued that the full boundary stress tensor is conserved in that ${\mathcal{D}}^{a} T_{ab} =0$.
We reconsider this relation in view of the fact that $\Delta^{ab} \neq 0$.

Recall that the full boundary stress tensor, $T_{ab}$, is
\begin{equation}
T_{ab} = T^{\pi}_{ab} - \frac{1}{8 \pi G} \Delta_{ab} .
\end{equation}
Expanding its divergence in a power series yields
\begin{align}
\mathcal{D}^a T_{ab} =& [\mathcal{D}^a T_{ab}]^{(0)} + \frac{1}{r^n} [\mathcal{D}^a T_{ab}]^{(1)} + \frac{1}{r^{n+1}} [\mathcal{D}^a T_{ab}]^{(2)} + \cdots \nonumber\\
=&  \bigg([\mathcal{D}^a T^{\pi}_{ab}]^{(0)} - \frac{1}{8 \pi G} [{\mathcal{D}^a} \Delta_{ab}]^{(0)} \bigg) + \frac{1}{r^n} \bigg([\mathcal{D}^a T^{\pi}_{ab}]^{(1)} - \frac{1}{8 \pi G} [{\mathcal{D}^a} \Delta_{ab}]^{(1)} \bigg) \nonumber\\
& + \frac{1}{r^{n+1}} \bigg([\mathcal{D}^a T^{\pi}_{ab}]^{(2)} - \frac{1}{8 \pi G} [{\mathcal{D}^a} \Delta_{ab}]^{(2)} \bigg) + \cdots \label{dvgcBStnsr}
\end{align}
where  $\mathcal{D}^a$ is the covariant derivative associated with $h_{ab}$.

For $n \geq 2$, we have
\begin{align}
T^{\pi}_{ab} =& - \frac{r}{8 \pi G} \bigg[ \frac{1}{r^n} \bigg( \frac{n}{2r^2} h^{(1)}_{ab} + \frac{1}{r^2} \gamma^{(1)} h^{(0)}_{ab} + \frac{n}{(n-1)} \alpha^{(1)} \mu^{(0)}_{ab} + \frac{1}{(n-1)} \gamma^{(1)} \mu^{(0)}_{ab} + \frac{1}{(n-1)} D_{a} D_{b} \alpha \bigg) \nonumber\\
& + \frac{1}{r^{n+1}} \bigg( \frac{n(n+1)}{2(n-1)r^2} h^{(2)}_{ab} + \frac{(n+1)(n+2)}{n(n-1)} \gamma^{(2)} \mu^{(0)}_{ab} - \frac{(n+1)}{(n-1)r^2} \gamma^{(2)} u^{(0)}_{a} u^{(0)}_{b} \bigg) + \cdots \bigg] ,
\end{align}
and plugging this and (\ref{Dltn0}) $-$ (\ref{Dltn2}) into (\ref{dvgcBStnsr}), we get
\begin{align}
\mathcal{D}^a T_{ab} &= \frac{1}{r^n} \bigg(\frac{1}{8 \pi G} \frac{1}{(n-1)r} D_{b} \gamma^{(1)} - \frac{1}{8 \pi G} \frac{1}{(n-1)r} D_{b} \gamma^{(1)} \bigg) + \frac{1}{r^{n+1}} \bigg( 0 - 0 \bigg) + \cdots, \nonumber\\
&= 0.
\end{align}
where $D^a$ is associated with $h^{(0)}_{ab}$. This verifies that the full boundary stress tensor, $T_{ab}$, is conserved.

For $n=1$, we take
\begin{align}
T^{\pi}_{ab} =& - \frac{r}{8 \pi G} \bigg[ \frac{1}{r} \bigg( -\frac{2}{r^2} \gamma^{(1)} u^{(0)}_{a} u^{(0)}_{b}  \bigg) + \frac{1}{r^2}\bigg( \frac{1}{2} \mu^{(2)}_{ab} + \frac{3}{2} \gamma^{(2)} \mu^{(0)}_{ab} - \frac{23}{2} \alpha^2 \mu^{(0)}_{ab}   \nonumber\\
& - 5 \alpha \gamma^{(1)} \mu^{(0)}_{ab} - (\gamma^{(1)})^2 \mu^{(0)}_{ab} - \frac{1}{2r^2} \mu^{(2)} u^{(0)a}u^{(0)b} + \frac{3}{r^2} \gamma^{(2)} u^{(0)a}u^{(0)b}  \nonumber\\
& + \frac{4}{r^2} \alpha \gamma^{(1)} u^{(0)a}u^{(0)b} + \frac{15}{r^2} \alpha^2 u^{(0)a}u^{(0)b} + \frac{4}{r^2} (\gamma^{(1)})^2 u^{(0)a}u^{(0)b} \bigg) + \cdots \bigg].
\end{align}
and substituting this with (\ref{Dlt12}) into (\ref{dvgcBStnsr}), it yields
\begin{align}
\mathcal{D}^a T_{ab} & = \frac{1}{r} \bigg( 0 - 0 \bigg) + \frac{1}{r^2} \bigg[\frac{1}{8 \pi G} \bigg( -  \frac{22}{r}  \alpha D_b \alpha -  \frac{3}{2 r}  \gamma^{(1)} D_{b} \alpha \bigg) -\frac{1}{8 \pi G} \bigg( \frac{3}{r}  D_{b} \gamma^{(2)} -  \frac{14}{r}  \alpha D_{b} \alpha -  \frac{20}{r}  \gamma^{(1)} D_{b} \alpha \bigg) \bigg]  + \cdots, \nonumber\\
&= \frac{1}{8 \pi G} \frac{1}{r^2}\bigg( - \frac{3}{r} D_{b} \gamma^{(2)} - \frac{8}{r} \alpha D_{b} \alpha + \frac{37}{2r} \gamma^{(1)} D_{b} \alpha \bigg)
\end{align}
where the second sub-leading order does not vanish. To be conserved, we require
\begin{equation}
\label{dvT12} D_b \bigg( 3 \gamma^{(2)} + 4 \alpha^2 - \frac{37}{2} \alpha \gamma^{(1)} \bigg) = 0
\end{equation}
and this relation can be satisfied by applying (\ref{cnstcmb11}) $-$ (\ref{cnstcmb13}).

%%%%%%%%%%%%%%%%%%%%%%%%%%%%%%%%%%%%%%%%%%%%
% Reference
%%%%%%%%%%%%%%%%%%%%%%%%%%%%%%%%%%%%%%%%%%%%

\bibliographystyle{plain}

\begin{thebibliography}{99}

\bibitem{Arnowitt:1960}
    R. Arnowitt, S.Deser and C. Misner, Phys.\ Rev.\  {\bf 117}, 1595 (1960); R.~L.~Arnowitt, S.~Deser and C.~W.~Misner, Phys.\ Rev.\  {\bf 118}, 1100 (1960) ; R.~L.~Arnowitt, S.~Deser and C.~W.~Misner, Phys.\ Rev.\  {\bf 122}, 997 (1961). ; R.~L.~Arnowitt, S.~Deser and C.~W.~Misner,  \textit{"The Dynamics of general relativity"}, [gr-qc/0405109]

\bibitem{Bondi:1962}
    H.~Bondi, M.~G.~J.~van der Burg and A.~W.~K.~Metzner,
    \textit{"Gravitational waves in general relativity. 7. Waves from axisymmetric isolated systems"}, Proc.\ Roy.\ Soc.\ Lond.\ A {\bf 269}, 21 (1962).

\bibitem{Geroch:1972}
    R.~P.~Geroch, \textit{"Structure of the gravitational field at spatial infinity"},  J.\ Math.\ Phys.\  {\bf 13}, 956 (1972).

\bibitem{Ashtekar:1978}
    A.~Ashtekar and R.~O.~Hansen, \textit{"A unified treatment of null and spatial infinity in general relativity. I - Universal structure, asymptotic symmetries, and conserved quantities at spatial infinity"}, J.\ Math.\ Phys.\  {\bf 19}, 1542 (1978).

\bibitem{Ashtekar:1979}
    A.~Ashtekar and A.~Magnon-Ashtekar, \textit{"Energy-Momentum in General Relativity"}, Phys.\ Rev.\ Lett.\ {\bf 43}, 181-184 (1979).

\bibitem{Ashtekar:1992}
   A.~Ashtekar and J.~D.~Romano, \textit{"Spatial infinity as a boundary of space-time"},  Class.\ Quant.\ Grav.\  {\bf 9}, 1069 (1992).

\bibitem{Penrose:1963}
    R. Penrose, \textit{"Asymptotic properties of fields and space-times"}, Phys.\ Rev.\ Lett.\  {\bf 10}, 66 (1963).


\bibitem{Brown:1992br}
  J.~D.~Brown and J.~W.~York, Jr.,\textit{"Quasilocal energy and conserved charges derived from the gravitational action"},
  Phys.\ Rev.\ D {\bf 47}, 1407 (1993)
  [gr-qc/9209012].
  %%CITATION = GR-QC/9209012;%%

\bibitem{Brown:1994gs}
  J.~D.~Brown, J.~Creighton and R.~B.~Mann,
  \textit{``Temperature, energy and heat capacity of asymptotically anti-de Sitter black holes,''}
  Phys.\ Rev.\ D {\bf 50}, 6394 (1994)
  [gr-qc/9405007].
  %%CITATION = GR-QC/9405007;%%

\bibitem{Hawking:1996}
    S.~W.~Hawking and G.~T.~Horowitz, \textit{"The Gravitational Hamiltonian, action, entropy and surface terms"},  Class.\ Quant.\ Grav.\  {\bf 13}, 1487 (1996)  [gr-qc/9501014].

\bibitem{MannRoss:1995}
R.~B. Mann and S.~F. Ross, {\it Cosmological production of charged black holes
  pairs},  {\em Phys. Rev. D} {\bf 52} (1995) 2254, [gr-qc/9504015].

\bibitem{Mann:2006}
    Robert B. Mann, and Donald Marolf, \textit{"Holographic Renormalization of Asymptotically Flat Spacetimes"},
    Class.\ Quant.\ Grav.\  {\bf 23}, 2927 (2006)  [hep-th/0511096].


\bibitem{Balasubramanian:1999re}
  V.~Balasubramanian and P.~Kraus,
  \textit{``A Stress tensor for Anti-de Sitter gravity,''}
  Commun.\ Math.\ Phys.\  {\bf 208}, 413 (1999)
  [hep-th/9902121].
  %%CITATION = HEP-TH/9902121;%%

  \bibitem{Mann:1999pc}
  R.~B.~Mann,
\textit{``Misner string entropy,'}
  Phys.\ Rev.\ D {\bf 60}, 104047 (1999)
  [hep-th/9903229].
  %%CITATION = HEP-TH/9903229;%%

  \bibitem{Emparan:1999pm}
  R.~Emparan, C.~V.~Johnson and R.~C.~Myers,
  \textit{``Surface terms as counterterms in the AdS / CFT correspondence,''}
  Phys.\ Rev.\ D {\bf 60}, 104001 (1999)
  [hep-th/9903238].
  %%CITATION = HEP-TH/9903238;%%


\bibitem{Mann:2008}
    Robert B. Mann, Donald Marolf, Robert McNees, and Amitabh Virmani, \textit{"On the Stress Tensor for Asymptotically Flat Gravity"}, Class.\ Quant.\ Grav.\  {\bf 25}, 225019 (2008)  [arXiv:0804.2079 [hep-th]].

\bibitem{Beig:1982}
    R. Beig and B. G. Schmidt, \textit{"Einstein's equations near spatial infinity"}, Comm.\ Math.\ Phys.\ {\bf 87}, 65-80 (1982).

\bibitem{Beig:1984}
    R. Beig, \textit{"Integration Of Einstein's Equations Near Spatial Infinity"}, Proc.\ R.\ Soc.\ Lond.\ A 391 295 (1984).

\bibitem{Astefanesei:2007}
    D.~Astefanesei, R.~B.~Mann and C.~Stelea, \textit{"Note on counterterms in asymptotically flat spacetimes"},  Phys.\ Rev.\ D {\bf 75}, 024007 (2007)  [hep-th/0608037].

\bibitem{Myers:1986}
    R. C. Myers and M. J. Perry, \textit{"Black Holes in Higher Dimensional Space-Times"}, Annals Phys.\  {\bf 172}, 304 (1986).

\end{thebibliography}

\end{document}